\documentclass[useAMS,usenatbib]{mn2e}
\usepackage{times}
\usepackage{amsmath}
\usepackage{amssymb,amsfonts}
\usepackage{graphicx}
\usepackage{verbatim}
\usepackage{psfig}

\def\ltsima{$\; \buildrel < \over \sim \;$}
\def\simlt{\lower.5ex\hbox{\ltsima}}
\def\gtsima{$\; \buildrel > \over \sim \;$}
\def\simgt{\lower.5ex\hbox{\gtsima}}
\def\kms{{\rm\,km\,s^{-1}}}

\def\kpc{{\rm\,kpc}}

\def\deg{^\circ}

\def\s{\ifmmode \widetilde \else \~\fi}
\def\={\overline}

\def\spose#1{\hbox to 0pt{#1\hss}}

\def\lta{\mathrel{\spose{\lower 3pt\hbox{$\mathchar"218$}}
     \raise 2.0pt\hbox{$\mathchar"13C$}}}
\def\gta{\mathrel{\spose{\lower 3pt\hbox{$\mathchar"218$}}
     \raise 2.0pt\hbox{$\mathchar"13E$}}}
\def\Dt{\spose{\raise 1.5ex\hbox{\hskip3pt$\mathchar"201$}}}    
\def\dt{\spose{\raise 1.0ex\hbox{\hskip2pt$\mathchar"201$}}}    

\def\dotsfill{\leaders\hbox to 1em{\hss.\hss}\hfill}

\title
[The AAT/WFI survey of the Monoceros Ring and Canis Major Dwarf galaxy]
{The AAT/WFI survey of the Monoceros Ring and Canis Major Dwarf galaxy: I. from {\it l} = (193 - 276)$^\circ$}
\author[Blair Conn et al.]
       {Blair C. Conn$^{1,2}$, Richard R. Lane$^1$, Geraint F. Lewis$^1$, Rodrigo Gil-Merino$^1$, Mike J. Irwin$^3$, 
       \newauthor Rodrigo A. Ibata$^4$, Nicolas F. Martin$^4$, Michele Bellazzini$^5$, Robert Sharp$^6$,
       \newauthor Artem V. Tuntsov$^1$ \& Annette M. N. Ferguson$^7$\\
$^{1}$Institute of Astronomy, School of Physics, A29, University of Sydney, NSW 2006, Australia\\
$^{2}$European Southern Observatory, Alonso de Cordova 3107, Vitacura, Santiago, Chile. Email:{\tt bconn@eso.org}\\
$^{3}$Institute of Astronomy, Madingley Road, Cambridge, CB3 0HA, U.K.\\
$^{4}$Observatoire de Strasbourg, 11, rue de l'Universit\'e, F-67000, Strasbourg, France\\
$^{5}$INAF - Osservatorio Astronomico di Bologna, Via Ranzani 1, 40127, Bologna, Italy\\
$^{6}$Anglo-Australian Observatory, Epping, NSW, 1710, Australia\\
$^{7}$Institute for Astronomy, University of Edinburgh, Royal Observatory, Blackford Hill, Edinburgh, EH9 3HJ, U.K.}

\begin{document}

\date{\today \hspace{10pt}(Version 2.6)} 

\pagerange{\pageref{firstpage}--\pageref{lastpage}} \pubyear{2006}

\def\LaTeX{L\kern-.36em\raise.3ex\hbox{a}\kern-.15em
    T\kern-.1667em\lower.7ex\hbox{E}\kern-.125emX}

\newtheorem{theorem}{Theorem}[section]

\label{firstpage}

\maketitle

\begin{abstract}
  We present  the results of an AAT  wide-field camera survey of  the stars in
  the Monoceros  Ring and  purported Canis Major  overdensity in  the Galactic
  longitudes of {\it  l} = (193 - 276)$^\circ$.  Current numerical simulations
  suggest that  both of these structures  are the result of  a single on-going
  accretion event,  although an  alternative solution is  that the  warped and
  flared  disc  of  the  Galaxy  can  explain the  origin  of  both  of  these
  structures.  Our results  show that,  with regards  the Monoceros  Ring, the
  warped and flared disc is unable to reproduce the locations and strengths of
  the  detections observed  around the  Galaxy. This  supports  a non-Galactic
  origin  for this  structure.  We report  8  new detections  and 2  tentative
  detections of  the Monoceros Ring  in this survey.  The exact nature  of the
  Canis Major  overdensity is still  unresolved although this  survey provides
  evidence that invoking  the Galactic warp is not  a sufficient solution when
  compared  with  observation.  Several  fields  in  this  survey  are  highly
  inconsistent with the  current Galactic disc models that  include a warp and
  flare, to such an extent that explaining their origins with these structures
  is problematic. We also report  that the Blue Plume stars previously invoked
  to support  the dwarf galaxy  hypothesis is unfounded, and  associating them
  with an  outer spiral arm  is equally problematic. Standard  Galactic models
  are  unable to  accommodate all  the observations  of these  new structures,
  leading away  from a  warped/flared disc explanation  for their  origins and
  more toward  a non-Galactic source.  Additionally, evidence is  presented in
  favour of  a detection of the Canis  Major dwarf stream away  from the Canis
  Major  region. As  the outer  reaches of  the Galactic  disc continue  to be
  probed, the  fascinating structures  that are the  Monoceros Ring  and Canis
  Major  overdensity  will  no doubt  continue  to  inform  us of  the  unique
  structure and formation of the Milky Way.
\end{abstract}

\begin{keywords}
Galaxy:\hspace{2pt}formation -- Galaxy:\hspace{2pt}structure -- galaxies:\hspace{2pt}interactions
\end{keywords}

\section{Introduction}
Recent deep optical/IR surveys, such as the Sloan Digital Sky Survey
(SDSS) \citep[]{2006ApJS..162...38A} and the Two-Micron All Sky Survey (2MASS)
\citep[]{2006AJ....131.1163S} are revealing increasingly complex structures in the Halo of the Milky
Way. Among these, several new dwarf galaxies and tidal streams have been uncovered,
but do these results favour the current galaxy formation paradigm? Can
we consider these new structures as supporting the $\Lambda$ Cold Dark Matter theories
\citep[][]{2003ApJ...591..499A,2003ApJ...597...21A} of galaxy
formation? Are these the low-mass satellites which will be gradually
accreted over time, and do they resolve previous issues such as the
missing satellite problem?

While it may be a little premature to answer these questions with the
current knowledge of Local Group environment, there is still
little direct evidence to support the idea of a merger history for the Milky Way. The Sagittarius dwarf galaxy
\citep{1994Natur.370..194I} has shown that such accretion does occur
and is in fact on-going but there are still too few nearby tidal stream
remnants to confirm it as the primary Galaxy formation mechanism.
However, analyses of these large surveys are uncovering new dwarf galaxies and tidal streams
in the inner halo of the Milky Way. For example, the discovery of tidal debris covering
60$\deg$ of the sky only 20 \kpc\ from the Sun, along with a new dwarf
galaxy and other tidal debris is presented by
\citet[][]{2006ApJ...639L..17G,2006ApJ...641L..37G,2006ApJ...643L..17G,2006ApJ...645L..37G}.
\citet[][]{2005AJ....129.2692W,2005ApJ...626L..85W}
have uncovered a new dwarf galaxy in Ursa Major along with the tiny Willman I
object, which is on the border between globular clusters and dwarf
galaxies. Even closer satellite galaxies have been found with \citet{Belo2006} and
\citet{2006ApJ...643L.103Z} presenting the discovery of another two dwarf
galaxies using the SDSS database. These two in Bo\"{o}tes and Canes
Venatici are located heliocentrically at $\sim$60 and
220\kpc respectively, and are, with the Ursa Major dwarf galaxy, pointing
to a possible resolution of the missing Galactic satellites problem
outlined in \citet{1999ApJ...522...82K}. Further structure is also
highlighted in the discussion by \citet{2006ApJ...637L..29B} on the presence
of tidal arms in the nearby globular cluster NGC5466.

Our nearest spiral neighbour, M31, also shows a complicated formation
history as \citet{2002AJ....124.1452F} have revealed in the Isaac Newton
Telescope Wide Field Camera Survey (INT/WFC).  M31 was long
thought to have been a relatively quiet spiral galaxy with a well
defined edge, but a faint diffuse outer disc, riddled with substructure, is now visible. Within
this substructure, a giant stellar stream \citep{2001Natur.412...49I}, a new
class of stellar cluster \citep{2005MNRAS.360.1007H} and the new
dwarf galaxy, Andromeda IX \citep{2004PASA...21..203L,2004ApJ...612L.121Z}, have been
discovered. As we probe our own Galaxy to similar depths, will we uncover
similar structure? The results coming from the SDSS and 2MASS
surveys are suggestive of this. The Milky
Way however, does seem to be a less complex system. Although
the Milky Way shows evidence of tidal debris in the Halo, within the Disc of
the Galaxy a major change in recent years has been the
discovery of the Monoceros Ring by \citet{2002ApJ...569..245N} and
its purported progenitor, the Canis Major dwarf galaxy
\citep{2004MNRAS.348...12M}. The alternative source of the excess
stars in Canis Major is the Argo Star System as discussed in \citet{2006ApJ...640L.147R}. 

The focus of the survey presented here has been to extend the INT/WFC
survey of the Monoceros Ring \citep{2005MNRAS.362..475C} around the
Galactic plane, as well as surveying around the Canis Major
region to provide insight into the possibility of locating a dwarf
galaxy there. The present survey has also attempted to investigate the
Triangulum-Andromedae overdensity \citep{2004ApJ...615..732R} and the
Virgo overdensity \citep{Juric2005}. These results will be presented elsewhere.

The layout of this paper is as follows: $\S$\ref{ring} summarises the discovery of the Monoceros Ring
and the Canis Major dwarf galaxy; $\S$\ref{obsaat} describes the
observational procedure and data reduction. Section~\ref{analysis} outlines the
method employed to analyse the data, the use of a synthetic Galaxy
model for comparison and the procedures for determining distances
and completeness. Section~\ref{Survey Fields} presents the data and
the discussion and conclusions of this study are
found in $\S$\ref{discussionaat} and $\S$\ref{conclusion}.
 
\section{The Monoceros Ring and the Canis Major dwarf galaxy}\label{ring}
Discovering the Monoceros Ring (MRi) via an overdensity of colour selected
F-stars in the SDSS dataset, \citet{2002ApJ...569..245N} described its original extent
from {\it l} = (170 - 220)$^\circ$. Subsequent surveys in 2MASS
\citep[][]{2003ApJ...594L.115R,2004MNRAS.348...12M} and the INT/WFC
\citep[][]{2003MNRAS.340L..21I,2005MNRAS.362..475C} extended the MRi
detections back towards the Galactic centre with some tentative
detections in the fields ({\it l,b}) = (61,$\pm$15)$^\circ$ and
(75,$+$15)$^\circ$. More recently, \citet{2006ApJ...637L..29B}, while
tracing the Sagittarius stream with SDSS, notes the presence of
the MRi in two bands at latitudes of {\it b} =
(20 - 30)$\deg$. Additionally, \citet{2006ApJ...651L..29G} discusses substructure in the
MRi, again revealed in the SDSS catalogue. Consistently, the MRi is found on both sides of the
plane of the Galaxy at galactocentric distances of $\sim$17 kpc. 

Revealing an overdensity in the 2MASS data, \citet{2004MNRAS.348...12M} fulfilled a prediction by
\citet{2002ApJ...569..245N} that a potential progenitor of the MRi
could lie in the nearby Canis Major constellation. Nestled
under the Galactic disc, this overdensity, dubbed the Canis Major
dwarf galaxy (CMa), can be found at ({\it l,b}) = (240,$-$9)$\deg$ and $\sim$7 \kpc\ from the Sun. Its close
proximity to the Galactic disc led \citet{2004A&A...421L..29M} to
argue that the overdensity was simply a consequence of the Galactic
warp. In response to this interpretation, \citet{2005MNRAS.362..906M}
presented results from a 2-degree Field (2dF) spectrographic radial velocity survey taken at the
Anglo-Australian Telescope. The initial results were complicated
through difficulties in removing the instrumental signature from the
data, but after resolving these issues \citet{2005MNRAS.362..906M}
maintained an interesting population of stars with a velocity
anomalous to the Galactic disc. Additionally, \citet{2005MNRAS.364L..13C}
showed that in the background of CMa, the MRi
was present at a distance of $\sim$13.5 \kpc, with a velocity of
$\sim$133 $\kms$ and a dispersion of 23 $\kms$. Finally, using radial
velocity data of the Carina dwarf and Andromeda galaxies, the MRi was revealed in
the foreground of these objects \citep{2006MNRAS.367L..69M}.  In front
of the Carina dwarf, it has properties of $<$V$_r$$>$ = 145$\pm$5 $\kms$ with a
velocity dispersion of only 17$\pm$5 $\kms$. In front of the Andromeda
galaxy, with stars taken from the ``One Ring'' field ({\it l,b}) =
(123,$-$19)$\deg$, it has properties of $<$V$_r$$>$ = $-$75$\pm$4 $\kms$ and a dispersion
of 26$\pm$3 $\kms$. Slowly, both a velocity and spatial distribution of the
MRi is being revealed.

The three main sources of evidence for the CMa dwarf are (in order of significance): the
overabundance of Red Clump and RGB stars in this region as seen in
the 2MASS catalogue \citep{2006MNRAS.366..865B}; the additional velocity component as seen in the
2dF survey of the CMa region and the presence of a strong Blue Plume population as
can be seen in Figure~1 of \citet{2005ApJ...633..205M}. Recently, the origins of the Blue Plume
population has been brought into question by
\citet{2005ApJ...630L.153C} and \citet{2006MNRAS.368L..77M}, who do not
associate these stars with the foreground overdensity, but rather part of a more distant population of
stars. This distant population is claimed to be the extension of the
Norma-Cygnus spiral arm into the CMa region; and so these stars now require
greater scrutiny in light of this new interpretation. \citet{2006A&A...451..515M} 
has also produced a more complete study of their warp scenario, in
which they confront the first and second sources of evidence
for CMa. They argue that not only can the CMa overdensity be explained by the warp, but that the velocity
dispersion as presented by \citet{2005MNRAS.362..906M} is expected by
current Galaxy models and shows nothing new. In studying the
outer disc they suggest the MRi is not a tidal arm from
an accretion event but rather the extension of the flare of the disc
into those latitudes. In short, they claim all of the new structures
in the disc of the Milky Way are simply explained in terms of known Galactic
structure. \citet{2006MNRAS.369.1911L} concludes on the ``Galactic warp versus
dwarf galaxy'' debate with the statement that the warp can be formed by such a wide variety of causes that neither
radial velocity or photographic surveys can disentangle
their origins. This is maybe the case, although a systematic survey of
the ages, metallicities, distances and abnormal velocity profiles in
this region should provide strong indications if a dwarf galaxy resides in the plane or not. The warp
is an important part of the puzzle but should not stop progress in
resolving this issue.

Although the results from this paper are unable to answer questions
regarding the velocity profile of these outer disc objects, it will
attempt to understand those of the CMa and MRi overdensities with
regard to their distance and position in the Galaxy. In our
discussion, we comment on the likelihood of the proposals put forward
by \citet{2006A&A...451..515M} in explaining the MRi in terms of the outer
disc and whether the warp satisfactorily describes the CMa overdensity.

\section{Observations and Reduction}\label{obsaat}
The data were obtained using the Anglo-Australian Telescope Wide Field Imager (AAT/WFI)
at Siding Spring Observatory (SSO) in New South Wales, Australia. Mounted at the telescope
prime  focus, the camera consisting of eight 4k$\times$2k  CCDs with
0.2295 arcsec per pixel, covers a field of view approximately 33'$\times$33' per pointing. 

The observations were taken over four observing runs, the first on
the 22$^{nd}$-25$^{th}$ January 2004, the second on 30$^{th}$ January - 1$^{st}$ February
2004, the third on the 14$^{th}$-16$^{th}$ of August 2004 and the fourth on the
1$^{st}$-5$^{th}$ February 2006. All of the fields were
observed with the $g$ (WFI SDSS \#90) and $r$ (WFI SDSS \#91) filters. These
were chosen to minimise the fringing effects that can be present when
observing with other filters. Each exposure consisted of a single 600
second exposure with the $g$ filter and two 450 second exposures with
the $r$ filter. Two exposures were performed in $r$ so as to
avoid preserving cosmic rays and overexposing the brighter stars when using a single 900
second exposure. Each night, twilight flats were taken along with
bias and dark frames for calibration and the closest Landolt Standard
Star field to our target was observed every two hours. In this
manner, the removal of instrumental signatures and precise photometric
calibration could be achieved. The seeing at the SSO can vary from
0.9$''$ to 3.0$''$ thus heavily affecting the limiting magnitude of the data. Although some fields were lost due to
poor weather conditions, only the fields with the best photometry have
been presented here. 

The present survey was designed as a continuation of the Monoceros Ring Survey observed with the Isaac Newton
Telescope Wide Field Camera \citep{2005MNRAS.362..475C}.  The fields were
chosen to be roughly separated by 20 degrees in Galactic Longitude
with adjustments made on the final location to ensure that the
field was placed to ensure minimal dust extinction. Altogether the present
survey has observed fields from {\it l} = (193 - 25)$^\circ$, across
the Galactic bulge; this paper reports only on the results of
those fields in the region {\it l} = (193 - 276)$\deg$.

\begin{figure}
\centerline{
\psfig{figure=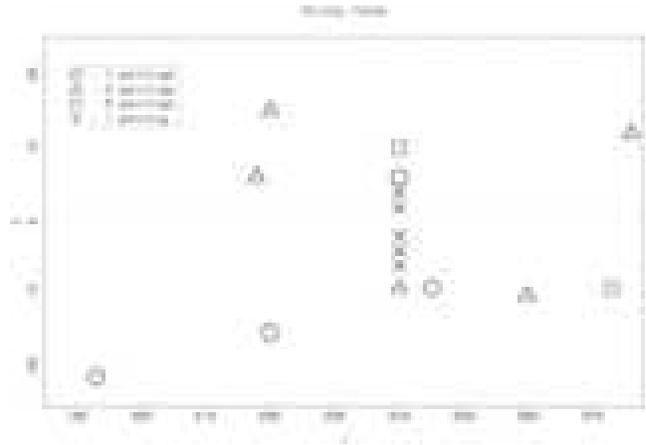,angle=270,width=\hsize}}
\caption[]{\small The location of the survey fields in the region
  ({\it l,b}) = (193 - 276)$\deg$. The symbols denote the number
  of pointings per field. The resultant fields have been selected based on data
  quality and while ideally four pointings would correspond to about 1
  square degree of sky, observations undertaken during February 2006
  had only 7 out of 8 CCDs available in the array.\label{figsurvey}}
\end{figure}

This part of the survey consists of 16 fields. Most fields are
approximately one square degree in size (four WFI pointings). However
some fields are a combination of one, two or three pointings depending
on the time available and quality of data obtained.  The single pointing fields make up a strip of observations
linking (240,$-$9)$^\circ$ with (240,$+$10)$^\circ$. A summary of the
field locations and area of the sky observed with the preliminary
results of this survey is shown in Figure~\ref{figsurvey} and Table~\ref{ObsTable}.

A specialised version of the CASU data reduction pipeline
\citep{2001NewAR..45..105I} was used to perform the de-biasing and
trimming, vignetting correction, astrometry  and photometry.   The
flat  fielding  of the science frames used a  master
twilight flat generated  over each entire observing run.  To account
for the dust extinction in the fields the  {\tt dust\_getval.c} program supplied by
Schlegel\footnote{http://www.astro.princeton.edu/$\sim$schlegel/dust/data/data.html}
was used to determine the extinction for each star. This program
interpolates the extinction from the dust maps of
\citet{1998ApJ...500..525S}. Using the several standard fields observed per
night provides a comparison for the calibration of the photometry to
be determined. The standards are used to derive the CCD zeropoints
from which all the magnitudes are determined \citep{2001NewAR..45..105I}.
A catalogue of each colour-band is produced by the pipeline for each
paired exposure of $g$ and $r$. Non-stellar images are rejected; however, near the
limiting magnitude, galaxies begin to appear stellar-like and thus
become a source of contamination in the dataset.

\begin{table*}
\centering
\caption{Summary of the observations  of Canis Major Tidal Stream with
  the AAT/WFI, ordered in ascending Galactic longitude ({\it l}). The
  variation between the total area calculations of 2004 and 2006 are
  due to failed CCDs in the array shrinking the field of view.}
\begin{tabular}{lcccccccl} \hline \hline
Fields  & Regions& Average Seeing& Total Area& Monoceros Ring & Average& Date \\ 
 ({\it l,b})$^\circ$& per field & (arcsec) &  ($deg^2$)  &  &  E(B-V)&  \\ \hline
 (193,$-$21)$^\circ$    & 4  & 1.3 & 1.21    & Maybe & 0.08 & 30/01/04\\
 (218,$+$6)$^\circ$     & 3  & 1.3 & 0.8     & Yes & 0.01&03/02/06\\
 (220,$-$15)$^\circ$    & 4  & 1.0 & 1.21    & Yes & 0.22&31/01/04\\
 (220,$+$15)$^\circ$    & 3  & 1.3 & 0.91    & Yes & 0.04&25/01/04\\
 (240,$-$9)$^\circ$     & 3  & 1.0 & 0.91    & Yes & 0.18&24/01/04\\
 (240,$-$6)$^\circ$     & 1  & 1.0 & 0.3     & No  & 0.40&31/01/04\\
 (240,$-$4)$^\circ$     & 1  & 1.0 & 0.3     & No  & 0.99&31/01/04\\
 (240,$-$2)$^\circ$     & 1  & 1.0 & 0.3     & No  & 1.10&31/01/04\\
 (240,$+$2)$^\circ$     & 1  & 1.0 & 0.3     & No  & 0.79&31/01/04\\
 (240,$+$4)$^\circ$     & 1  & 1.0 & 0.3     & Yes & 0.28&31/01/04\\
 (240,$+$6)$^\circ$     & 4  & 1.2 & 1.21    & Yes & 0.13&31/01/04\\
 (240,$+$10)$^\circ$    & 2  & 1.3 & 0.61    & Yes & 0.10&01/02/04\\
 (245,$-$9)$^\circ$     & 4  & 1.6 & 1.21    & Yes & 0.14&30/01/04\\
 (260,$-$10)$^\circ$    & 3  & 1.3 & 0.91    & No  & 0.21&01/02/04\\
 (273,$-$9)$^\circ$     & 2  & 1.4 & 0.53    & No  & 0.29&03/02/06\\
 (276,$+$12)$^\circ$    & 3  & 1.1 & 0.8    & Yes & 0.09&01/02/06\\
\hline\hline
\end{tabular}
\label{ObsTable}
\end{table*}

\section{Data Preparation}\label{analysis}
\subsection{Detecting Non-Galactic Structure}\label{detections}

\begin{figure}
\centerline{
\psfig{figure=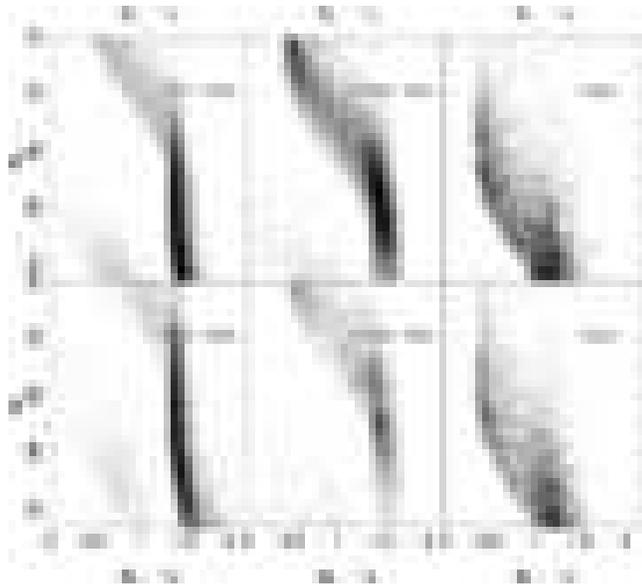,angle=0,width=\hsize}}
\caption[]{\small The top panels show the Hess plots (a pixelated Colour
  Magnitude diagram in which the resultant image is the square root of
  the number density) of (240,$-$2)$^\circ$ field from the
  synthetic model of the Galaxy by \citet{2003A&A...409..523R} being split into its various
  Galactic components - Thin Disc, Thick Disc and Halo. This illustrates where the various
  components of the galaxy lie on the CMD (Colour-Magnitude Diagram).  The lower panels are from the model
  field (193,$-$21)$^\circ$ and show how the components vary further away from the plane of the Galaxy.
\label{figmodel}}
\end{figure}

\begin{figure}
\centering
\includegraphics[width=\hsize]{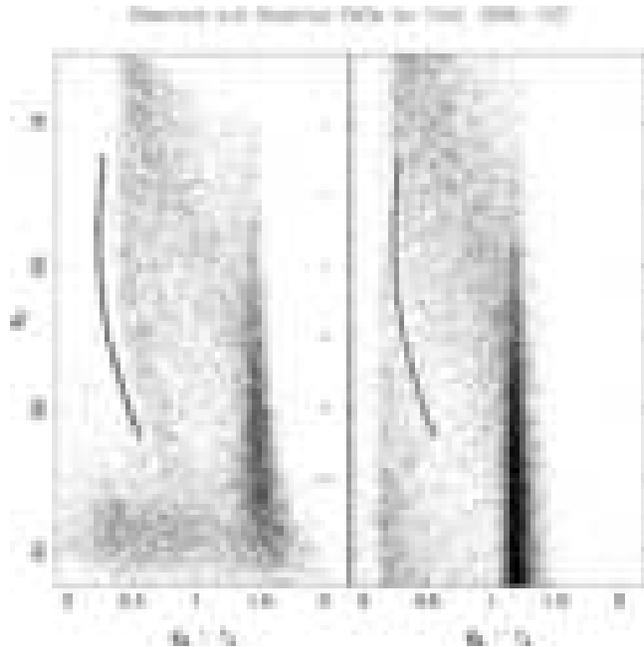}
\caption[]{\small On the left is the AAT/WFI CMD of the field ({\it l,b}) =
  (220,$+$15)$\deg$, and on the right is the same field as produced by
  the Besan\c{c}on synthetic Galaxy model. The model outputs are
  CFHTLS-Megacam (AB) Photometric system converted to Sloan
  $g'$ and $r'$ using Equation~\ref{CFHTtoSDSS}. The fiducial sequence is placed with zero offset and uses the raw
  SDSS data \citep{2002ApJ...569..245N} from which the fiducial was created.
\label{figsloancomp}}
\end{figure}

\begin{figure}
\centerline{
\psfig{figure=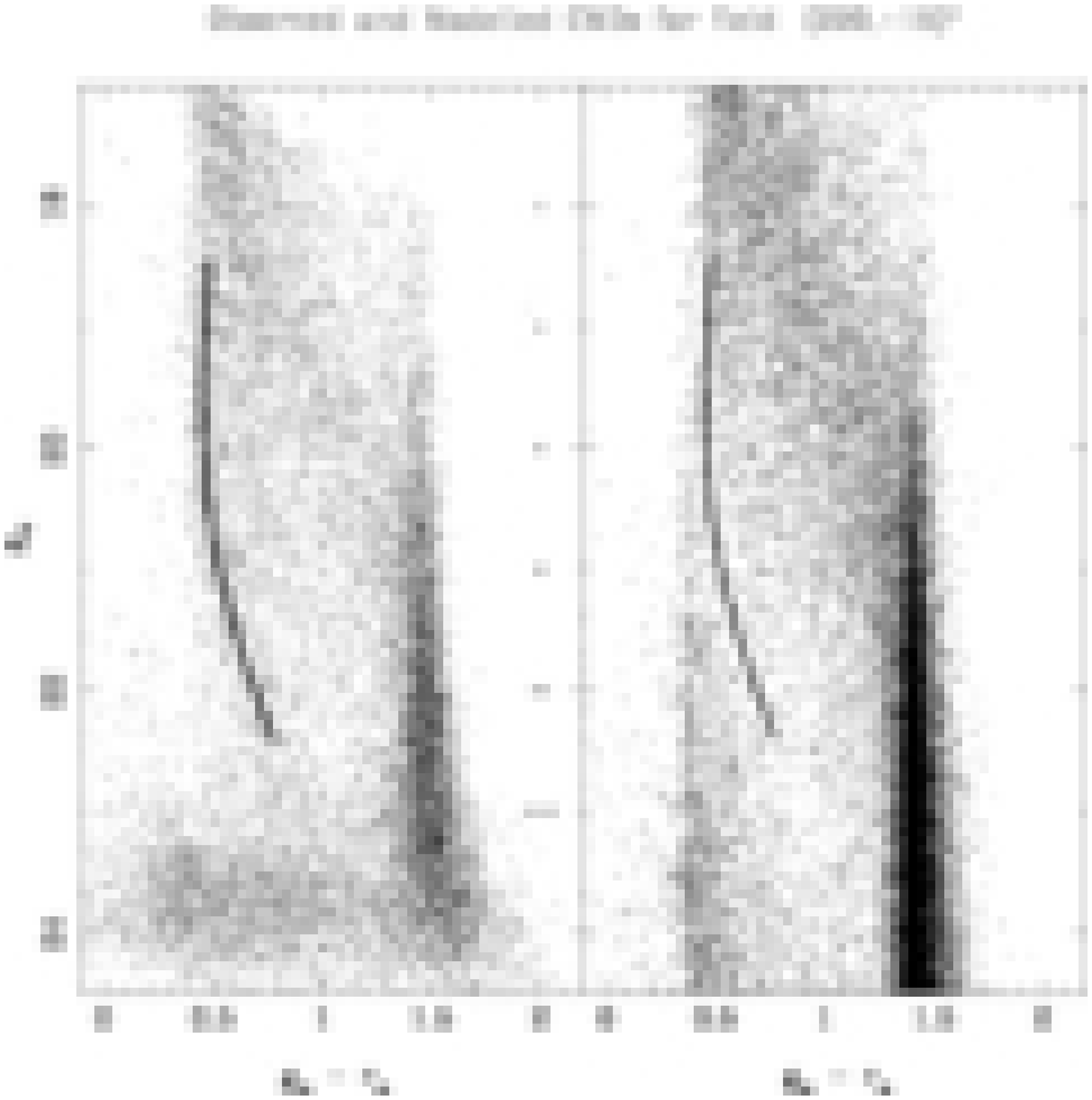,angle=0,width=\hsize}}
\caption[]{\small On the left is the AAT/WFI CMD of the field ({\it l,b}) =
  (220,$+$15)$\deg$, and on the right is the same field as produced by
  the Besan\c{c}on synthetic Galaxy model. The model is now the Sloan
  $g'$ and $r'$ in the AB system converted to the INT/WFC $g$ and $r$
  in the Vega Photometric system via Equation~\ref{colourconversion0}. The fiducial
  sequence has also undergone the same colour conversion.
\label{figvegacomp}}
\end{figure}

Searching for additional structure within the Colour-Magnitude
diagrams (CMDs) of the AAT/WFI survey
requires an understanding of the inherent structures
present when observing through the Galactic
disc. The Milky Way can be roughly divided into four components: Thin
disc, Thick disc, Halo and Bulge. The Thin disc is essentially the
plane of the Galaxy containing the majority of the stars and is where the spiral
structure is also present. The Thick disc is less dense than the Thin disc and has a greater scale height out of
the plane. The Halo is considered to be a smooth spherical distribution of
stars around the centre of gravity of the Galaxy and extends out
beyond 20 - 30 kpc. The Bulge is the central region of the Galaxy, including the Bar,
and contains a very dense old stellar population. The Besan\c{c}on synthetic Galaxy
model\footnote{http://www.obs-besancon.fr/www/modele} \citep{2003A&A...409..523R} allows for these components to be
considered separately as shown in Figure~\ref{figmodel}. Each component is revealed in a 
distinct region of the CMD, allowing for at least a preliminary
estimate of its origins when interpreting the CMDs from the
observational data. It should be noted that the Besan\c{c}on model
employs a Thin disc cut-off at $\sim$14 \kpc\ from the Galactic
centre. The validity of such a cut-off is disputed in \citet{2006MNRAS.369.1911L}.
 
As an example, Figure~\ref{figmodel} shows the breakdown of the Galactic components
as seen in two one square degree fields at (240,$-$2)$\deg$ and
(193,$-$21)$\deg$. Analysing Figure~\ref{figmodel} reveals that, as
expected, fields closer to the plane have more stars than those away
from the plane. This manifests in the smoother CMDs in the top
panels, with the bottom panels containing fewer Thin and Thick disc stars. The structure of each component is due to the
homogeneity of the Galaxy and as a result a Main Sequence forms at
every distance increment along that line of sight. So as the Milky Way is probed
deeper, the collective weight of the Main Sequences sum to form the
CMDs that are observed. The closest Main Sequences, being brighter
in apparent magnitude, form at the top of the CMDs with the
subsequent Main Sequences, being fainter, forming lower down on the
CMDs. The different stellar components of the Thin and Thick disc
shows the significant change in density between the two, giving them
different characteristics in the CMDs. The Thin disc being narrow and
dense has a smaller scale height (typically estimated at several
hundred parsecs thick) which forms a much tighter sequence in the CMD. The
more diffuse Thick disc being more extended out of the plane (several
kiloparsecs, thick) forms a broader sequence. Comparing the top and
bottom panels in Figure~\ref{figmodel} also shows the importance of
the angle of the observations with regard the Galactic Plane.

Detecting non-Galactic components then involves visually comparing
the strong sequences detected in the observations with what is
expected from the models. The Monoceros Ring is located beyond the
edge of the Thick disc, and thus should be seen as a coherent sequence
below the last sequence of the Thick disc, as can be seen in Figure 12
of \citet{2002ApJ...569..245N}. Differentiating it from the Halo
component is simpler due to the Halo not having any strong Main Sequences
present below the Thick disc in the range of our CMDs. So, any strong
sequence below the Thick disc will most likely be of non-Galactic
origins. Although given the recent hypothesis on the MRi being
attributed to the flare of the disc, this will also need to be considered. Another
possibility is that a non-spherical halo could be misinterpreted as a
``non-Galactic'' feature in the outer Thick disc. Comparisons with the
Besan\c{c}on synthetic galaxy model reveals that their spheroid
density distribution has a flattening of 0.76 and a power index of 2.44
\citep{2000A&A...359..103R} which is moderately non-spherical. They note a significant degeneracy
between these parameters which could allow for spheroids with a (c/a) of
0.85 (close to spherical) or 0.6 (quite oblate). As discussed in the following sections, the synthetic
galaxy model does not introduce structures into the outer Thick disc
and so the influence of a non-spherical halo can be assumed to be
minimal. The existence of the Canis Major dwarf galaxy is currently being scrutinised and so the
CMDs of the CMa region will be used to investigate the influence of
the Galactic warp on the stellar populations present there and whether
it is a viable solution to resolving the dwarf galaxy debate. 

The synthetic CMDs from the Besan\c{c}on model data are
considered out to 100 kpc (heliocentric distance, R$_{HC}$) which
ensures that the Halo population extends below the magnitude cut-off of
the data and hence does not introduce additional structure. The
synthetic fields contain a broad magnitude range ($-$99,99) for each
passband with zero extinction and the inclusion of all ages/populations
and luminosity classes.  This then provides as complete a picture of
the region of interest as the model can supply. It also allows us to
apply colour and distance cuts at our discretion.  The model can
output directly to the bands $g_{CFHT}$ \& $r_{CFHT}$ in the
CFHTLS-Megacam
(AB)\footnote{http://www.cfht.hawaii.edu/Instruments/Imaging/MegaPrime/}
photometric system which need to be converted to the SDSS ($g'$,$r'$)
system via Equation~\ref{CFHTtoSDSS}.

\begin{equation}
\begin{aligned}
  r' &= r_{CFHT}\label{CFHTtoSDSS}\\
  g' &= \frac{g_{CFHT} - 0.1480\times r_{CFHT}}{1 - 0.1480}\\
\end{aligned}
\end{equation}

\citet{1994AJ....108.1476F} report that no conversion is needed
between SDSS passbands ($g'$,$r'$) in the AB system to SDSS ($g'$,$r'$) in the
Vega system. However, Figure~\ref{figsloancomp} shows that when using only AB
magnitudes, the AAT/WFI data in the Vega system does not match the model in
the AB system. A fiducial of the ridgeline from the original SDSS detection of
the Monoceros Ring by \citet{2002ApJ...569..245N} can be seen to lie blueward of the bluest edge
of the data (See $\S$~\ref{distance}). So, using the results of \citet{2003MNRAS.340L..21I}, who
compared overlapping INT and SDSS fields to determine a colour
conversion between the two systems, the model is now shifted to the
Vega system in this way.  This is applicable to our data as
both the INT dataset and our AAT dataset have been reduced using the
same pipeline (with small modifications to allow for differences
between the telescopes) and no such study has been undertaken with the
AAT/WFI instrument. The resultant colour transformation to the INT
($g$,$r$) is:


\begin{equation}\label{colourconversion0}
  \begin{aligned}
    &  (g-r) = 0.21 + 0.86\times(g'-r')\\
    &  g = g' + 0.15 - 0.16\times(g-r)
    \end{aligned}
\end{equation}

The effect of this transformation can be seen in
Figure~\ref{figvegacomp}. The converted SDSS fiducial, corrected via Equation~\ref{colourconversion0}, is now an excellent
fit to the strong sequence in these data. The photometric system of
the model CMD is now the same as the AAT/WFI CMD and no further changes have been applied to the
model or data. All figures in this paper will be shown in the same format as Figure~\ref{figvegacomp}, utilising
Equations~\ref{CFHTtoSDSS} and \ref{colourconversion0}, with the
resultant ($g,r$) from the synthetic Galaxy model considered the same as
the extinction corrected observational dataset
($g_{\circ}$,$r_{\circ}$). The Besan\c{c}on synthetic Galaxy model
employs different density profiles for each component of the Galaxy. These have been outlined
in \citet{2005MNRAS.362..475C} and in more detail in
\citet[][]{1986A&A...157...71R,1996A&A...305..125R,2000A&A...359..103R}
and \citet{2001A&A...373..886R}.

\subsubsection{Magnitude Completeness}\label{completeness}
Most of the fields presented here, consist of several overlapping
subfields, see column 2, Table~\ref{ObsTable}. The completeness of
this sample is determined in a similar manner to that of the 2MASS
All-Sky Point Source Catalogue \citep{2006AJ....131.1163S}.  This
approach determines the fraction of stars that are detected in both
overlapping images as a function of magnitude. So by matching the
stars within those overlapping regions and calculating the proportion
of matched stars in each magnitude bin with respect to the total number of
stars observed, an estimate of the completeness is made. This produces a photometric completeness curve which
can fit approximately by the equation:
\begin{equation}\label{eqncomplete}
  CF = \frac{1}{1 + e^{(m - m_c)/ \lambda}}\\
\end{equation}
where m is the magnitude of the star, m$_c$ is the magnitude at 50\%
completeness and $\lambda$ is the width of the rollover from 100\% to
0\% completeness. The values used to model each field can be
found in Table~\ref{CompTable}.   
\begin{table}
\centering
\caption{\small Parameters used to model the completeness of each field,
  ordered in ascending Galactic longitude ({\it l}). $m_c$ is the
  estimated 50\% completeness level for each filter with $\lambda$ describing the
  width of the rollover function (Equation~\ref{eqncomplete}). }
\begin{tabular}{lccc}
\hline
Fields ({\it l,b}) & $m_c$ ($g_\circ$) & $m_c$ ($r_\circ$) & $\lambda$\\
\hline
(193,$-$21)$^\circ$ & 22.40 & 21.80 & 0.40\\
(218,$+$6)$^\circ$ & 23.40 & 22.40 & 0.50\\
(220,$-$15)$^\circ$ & 22.90 & 21.90 & 0.30\\
(220,$+$15)$^\circ$ & 23.60 & 22.60 & 0.30\\
(240,$-$9)$^\circ$ & 23.60 & 22.50 & 0.40\\
(240,$+$6)$^\circ$ & 24.00 & 22.70 & 0.40\\
(240,$+$10)$^\circ$ & 22.65 & 21.70 & 0.25\\
(245,$-$9)$^\circ$ & 23.30 & 22.20 & 0.60\\
(260,$-$10)$^\circ$ & 22.80 & 21.80 & 0.40\\
(273,$-$9)$^\circ$ & 23.40 & 22.20 & 0.50\\
(276,$+$12)$^\circ$ & 23.70 & 22.50 & 0.40\\
\hline
\end{tabular}
\label{CompTable}
\end{table}

Although the completeness of our survey is not a key problem, attempting to
characterise it does provide a manner in which we can compare the data
quality of the various fields. In general, this allows an
estimate of the magnitude at which the data becomes
untrustworthy. Additionally, since the model is mostly used to help identify the major structures
in the CMDs, it is unnecessary to apply the completeness function to the
model. This is because those structures are typically well away from
the 50$\%$ completeness limit. It is also important to note that while this method does not account for stars in crowded
fields, none of our survey fields can be considered crowded and so
this approach is valid for the entire dataset.

\subsubsection{Estimating the Distance and Additional
  Calibration}\label{distance}

Determining the distance to the Canis Major and Monoceros Ring
sequences is achieved by taking the ridge-line of the detection in the
SDSS S223$+$20 field [\citet{2002ApJ...569..245N}, Figure 12] and
creating a fiducial sequence. The AAT/WFI uses SDSS filters
and so the fiducial sequence can be directly used on the data
with the colour transformations described in Equation~\ref{colourconversion0}.

The heliocentric distance estimate of the SDSS S223$+$20 detection is
assumed to be 11.0 \kpc\ \citep{2002ApJ...569..245N}, this then is the zero offset
distance.  Since the fields have been extinction corrected it is assumed that only distance variations are the cause
for any deviation in magnitude from this location. This method does not
account for possible differences in age or metallicity between Main Sequences. The heliocentric distance is calculated
using Equation~\ref{helio} and assuming the Sun is 8.0 \kpc\ from the
Galactic centre, the galactocentric distance is found from simple trigonometry.

\begin{equation}\label{helio}
  R_{HC} = 11.0 \times\biggl(10^\frac{offset}{5.0}\biggr)
\end{equation}

Determining a value for the error associated with such a measurement is dependent on several factors.
Most predominant of these is whether the fields have been correctly
calibrated with regard to their photometry and taking into account the
dust extinction present within the fields. The
dust correction for this data will always over-correct for stars
within the Galaxy, because the dust value is based on the entire
cumulative impact of the dust along that line-of-sight \citep{1998ApJ...500..525S}. The stars in
this survey do not reside at the end of that line-of-sight and so will
be over-corrected in the dust extinction process. In most of the
fields, the levels of dust are sufficiently low that the difference
between the dust value used and the ``correct'' value should be
small, see Table~\ref{ObsTable}. 

To determine whether the colour transformations applied to the data set
correspond to reliable distance estimates, three fields to which there
are distance estimates to these structures from other surveys have been analysed. Those are ({\it l,b}) =
(220,$+$15)$\deg$, (240,$-$9)$\deg$ and (245,$-$9)$\deg$. The first field is very close to
the original Monoceros Ring detection of \citet{2002ApJ...569..245N}
and the second and third are part of the Canis Major detection fields of
\citet{2005MNRAS.362..906M}. The MRi is known to have a distance of
11.0 \kpc\ in the \citet{2002ApJ...569..245N} field, at
(220,$+$15)$\deg$ it is also located at 11.0 \kpc. For the CMa fields, the distance determined by
\citet{2005MNRAS.362..906M} is about 7.2 \kpc. The present survey
locates the centre of the strong sequence at 7.3 \kpc\ or $-$0.9
magnitudes of offset. Importantly then, each distance estimate here is
consistent with independent measurements of those structures.

Having understood the errors involved in both the determination of the
photometry, extinction correction and the fiducial sequence,
manually placing this fiducial at the centre and two extremes of an acceptable
fit ``by-eye'' provides a range of distances over which this structure
resides. Given the large errors involved, these distances can only be
considered indicative of the true distance. However, the range of
magnitude offsets defining the extremes do give a sense of the width
of the structure. Several fields have only one distance estimate, as
locating the extremes is not possible due to the data quality or the
narrowness of the sequence. The dominant strong sequence in each
field, which could be attributed to either the Thin/Thick disc or CMa
overdensity has only a single distance estimate, corresponding to the
faintest edge of that feature. This is due, in general, to the lack of
a definite upper edge, see Table~\ref{ResultsTable}.

\subsubsection{Signal to Noise Estimation Technique}\label{snrest}
Estimating the signal to noise ratio (S/N) of the Monoceros Ring Main
Sequence found in the data has been attempted for several fields. The
criteria for S/N determination is that the MRi Main Sequence is
sufficiently distinct from the CMa/Disc sequence to avoid potential
contamination. The model field is assumed to represent the properties
of the background Galactic stars which should be removed to highlight
the additional MRi Main Sequence. The model needs to be adjusted first
to more accurately reflect the properties of the data and then
subtracted to reveal the excess MRi stars. Before subtraction the
model is degraded to match both the completeness
profile of the data as shown in Table~\ref{CompTable} and the relevant
magnitude error properties. To ensure that similar numbers of
stars are present in both the data and model prior to
subtraction an additional scaling factor is applied. These processes result in
the data and the model being essentially
identical with the exclusion of the additional MRi Main Sequence. The
S/N is estimated by dividing the number of stars in the feature by the
Poisson noise due to the modelled number of stars in the region\footnote{Parameters
  available on request: bconn@eso.org}. Only five fields have
been suitable for this estimate namely ({\it l,b})$\deg$ =
(218,$+$6)$\deg$, (220,$+$15)$\deg$, (220,$-$15)$\deg$, (240,$+$10)$\deg$
and (276,$+$12)$\deg$, see Table~\ref{ResultsTable}. These fields have
MRi detections which are easily distinguished from the dominant Main
Sequences in the CMD and thus are suitable for this technique. The
remaining detections are too close to the CMa/Disc population to
easily measure their S/N ratio with this method.

\section{Survey Fields}\label{Survey Fields}

\begin{table*}
\centering
\caption{\small Summary of the observations  of Canis Major Tidal Stream with
  the AAT/WFI, ordered in ascending Galactic longitude ({\it l}). }
\begin{tabular}{lcccccc} \hline \hline
Fields ({\it l,b})$^\circ$ & MRi offset  & MRi dist & MRi S/N & MW/CMa offset & MW/CMa dist \\
 & \& width (mag) & \& width (kpc) & estimate &Lower edge (mag)  & (kpc) \\ \hline
 (193,$-$21)$\deg$ & 0.5                 & 13.8    &     -           & $-$1.2 & 6.3\\
 (218,$+$6)$\deg$  & 0.0$^{+0.35}_{-0.3}$& 11.0$^{+1.9}_{-1.4}$&  $\sim$34  & $-$1.5 & 5.5\\
 (220,$-$15)$\deg$ & 0.2$^{+0.3}_{-0.3}$ & 12.1$^{+1.7}_{-1.6}$ & $\sim$32  & $-$1.4 & 5.8\\
 (220,$+$15)$\deg$ & 0.0$^{+0.3}_{-0.35}$ & 11.0$^{+1.6}_{-1.6}$ & $\sim$14 & $-$2.5 & 3.5\\	
 (240,$-$9)$\deg$  & 0.4                & 13.2   &  -      & $-$0.6 & 8.3 \\
 (240,$-$6)$\deg$  & -                   & -     &   -      & -    & - \\
 (240,$-$4)$\deg$  & -   & -  & - & - & - \\
 (240,$-$2)$\deg$  & -   & -  & - & - & - \\
 (240,$+$2)$\deg$  & -   & -  & - & - & - \\
 (240,$+$4)$\deg$  & 0.1                 & 11.5  & -         & $-$1.4 & 5.8 \\
 (240,$+$6)$\deg$  & 0.3$^{+0.3}_{-0.3}$ & 12.6$^{+1.9}_{-1.6}$& - & $-$1.2 & 6.3 \\
 (240,$+$10)$\deg$ & 0.5$^{+0.3}_{-0.3}$ & 13.8$^{+2.1}_{-1.7}$& $\sim$22 & $-$2.2 & 4.0 \\
 (245,$-$9)$\deg$  & 0.5                 & 13.8  & -         & $-$0.6 & 8.3 \\
 (260,$-$10)$\deg$ & -                   &  -    & -        & $-$0.3 & 9.6 \\
 (273,$-$9)$\deg$  & -                   &  -    & -        & $-$0.3 & 9.6 \\
 (276,$+$12)$\deg$ & 0.3                 & 12.6  & $\sim$26        & $-$2.0 & 4.4 \\
\hline\hline
\end{tabular}
\label{ResultsTable}
\end{table*}

\begin{figure}\centering
\includegraphics[width=\hsize,angle=0]{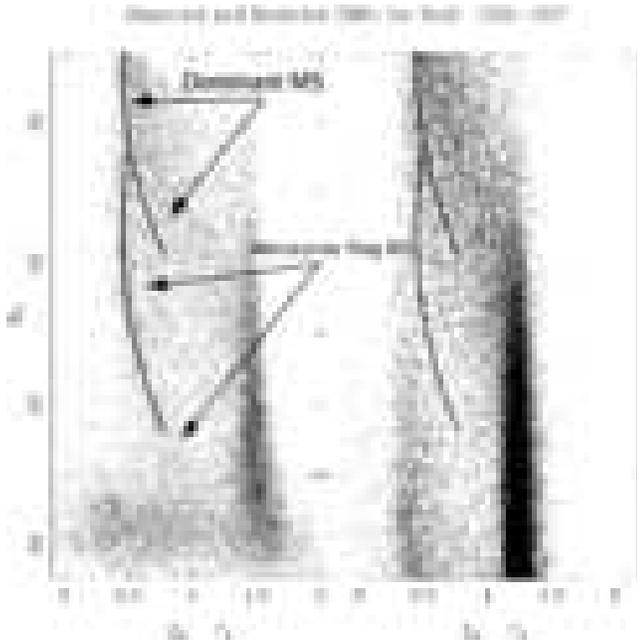}
\caption[]{\small Hess plot (a pixelated Colour-Magnitude diagram where the grayscale is the square root of
  the number density for that pixel) of  ({\it l},{\it b})$\deg$ = (220,$+$15)$^\circ$ (Left) and the
  corresponding Besan\c{c}on model (Right). This figure illustrates
  the alignment of the fiducial with both the dominant Main Sequence
  in the CMD associated with brighter nearby stars and the fiducial
  aligned with the MRi feature. As illustrated, the dominant Main
  Sequence is only defined by its fainter edge while the MRi feature
  is located along the centre of the Main Sequence. For this feature
  the offsets employed are $-$2.5 and 0.0 mag. To
  avoid unnecessarily cluttering the data CMDs, the following figures
  only show the placement of the fiducial on the model. This allows a
  direct comparison with model and highlights where the new features
  lie with respect to the expected Galactic components.
\label{fig220p15example}}
\end{figure}

The location of each field is shown graphically in
Figure~\ref{figsurvey}.  Each field presented in this section shows
the CMDs with the appropriate fiducial sequence taken from the
original \citet{2002ApJ...569..245N} detection as described in the
previous section. The CMDs that we have used are density maps of the
underlying distribution with the square root of the number of stars
per pixel being presented in all of the CMDs in this paper, called a
Hess diagram. This method provides better contrast of the structures
especially in high star density regions. In the following sections, the distance estimates to the major features present in
each CMD are provided with an analysis of these results in the
Discussion section. The principal features of the CMD have been identified by
visual comparison with the Besan\c{c}on model. The MRi Main Sequence
is evident by its absence in the model and the CMa Main Sequence is
inferred to be associated with the strongest Main Sequence in the CMD. The fiducial
is then placed on the CMD and shifted in magnitude to be aligned
``by-eye'' with the respective features. A more accurate technique is
unneccessary given the inherent uncertainties already present in a
distance estimate of this kind. It is important to note that the
fiducials placed on the dominant features of the CMD are not
necessarily considered detections, but rather placed to provide
insight into all the features present in the CMDs. For the dominant
and strongest sequence in each CMD, only the fainter edge has been fit
with the fiducial sequence providing a single distance estimate. For
the fainter coherent sequences, both the upper and lower extremes have
been fitted giving a range of distances to that structure, see Figure~\ref{fig220p15example}. The
complete list of the magnitude offsets and distance estimates is
contained in Table~\ref{ResultsTable}. The brighter, nearer or more
dominant sequence has been listed under the heading MW/CMa (Milky
Way/Canis Major) to illustrate that differentiating between the two is
not straightforward when using CMDs. Although some fields are most
likely to contain only Milky Way stars, there are several that are
possibly a mix, or completely dominated by CMa stars. For this reason,
this structure is left ambiguously identified in
Table~\ref{ResultsTable}. The fainter, more distant or less dominant
sequence is listed under the heading MRi. To understand the widths of
the structures and hence an estimation of the errors for each field,
this table should be referenced. The final structure of interest are
the Blue Plume stars which are located around 18$^{th}$
magnitude and in the colour range ($g-r$) = 0.0 - 0.3. These stars can
be seen clearly in Figure~\ref{fig240m9} and are discussed further in $\S$~\ref{BPstars}.

\subsection{Monoceros Region}
The four fields presented here are in close proximity to the original
detections of the Monoceros Ring as presented in
\citet{2002ApJ...569..245N}. Importantly, the present survey has sampled
both sides of the plane, finding the MRi to be present across the
Galactic plane. This has implications regarding whether the MRi could
be a phenomena related to the Galactic warp and flare.

\subsubsection{Fields at \bf$(193,-21)\deg$}\label{195m20des}
\begin{figure}
\centering
\includegraphics[width=\hsize,angle=0]{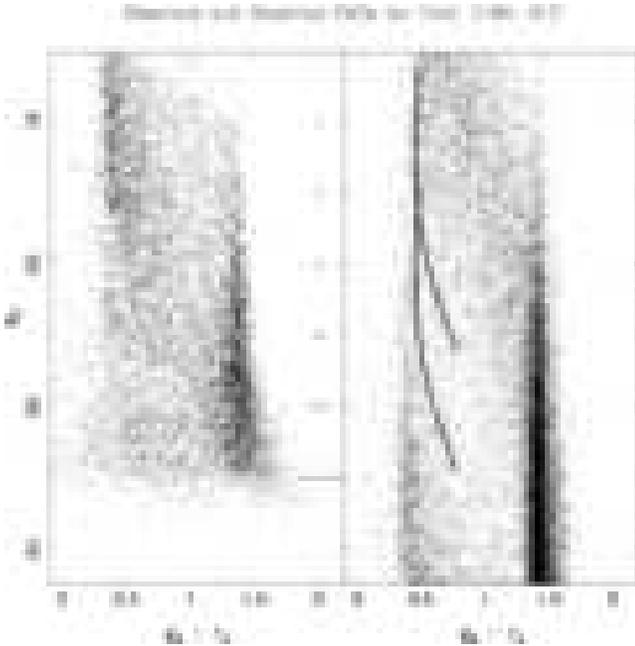}
\caption[]{\small Hess plot (a pixelated Colour-Magnitude diagram where the grayscale is the square root of
  the number density for that pixel) of ({\it l},{\it b}) =
  (193,$-$21)$^\circ$ (Left) and the corresponding Besan\c{c}on model (Right). The synthetic Galaxy model is taken from
  the Besan\c{c}on online galaxy model website. The model
  has a distance interval of 100 \kpc\ from the sun to ensure no
  artificial cuts enter into the CMDs. The model is selected
  in $U$ band and then converted to $g,r$ using colour corrections
  from the INT/WFC website. There are two structures in this field, the strong sequence
  beginning at $g_\circ$$\sim$19.0 and a fainter coherent sequence at
  about 20$^{th}$-21$^{st}$ magnitude. The offsets required to fit the
  fiducial sequence from \citet{2002ApJ...569..245N} are $-$1.2 mag and 0.5
  mag. Heliocentrically, this corresponds to 6.3 and 13.8 \kpc. The 50\%
  completeness in $g_\circ$ for this CMD is 22.4 mag.
\label{fig195m20}}
\end{figure}

This is the furthest field West from Canis Major (towards
the anti-centre direction) that is included in this survey (Figure~\ref{fig195m20}). Despite the good seeing
and area covered in this field, many of the fainter stars have been
lost due to the relative proximity of the Moon. This has removed a lot
of the detail present in other fields of similar size. Using the
fiducial sequence on the two main features in this CMD, we obtain two
distance estimates. The first using an offset of $-$1.2 mag
corresponding to a heliocentric distance of $\sim$6.3 \kpc. The second
more distant feature, somewhat more tentative, is found at an offset
of 0.5 mag, or R$_{HC}$ $\sim$ 13.8 \kpc. Given the lack of
clarity regarding the potential Monoceros Ring feature in the data, no
attempt has been made to measure the spread of distances over which it
is visible. Indeed, it is uncertain whether this is simply the strong
overdensity of stars seen at the faint blue end of the model CMD. The
stars located at the faint blue end of the CMD are most likely
misclassified galaxies and are unlikely to represent any real Galactic
structure, they have a considerable error in colour as can be seen by
the error bars on the right hand side of the panel. The 50\%
completeness in $g_\circ$ for this CMD is 22.4 mag.

\subsubsection{Fields at \bf$(218,+6)\deg$}\label{218p6des}
\begin{figure}\centering
\includegraphics[width=\hsize,angle=0]{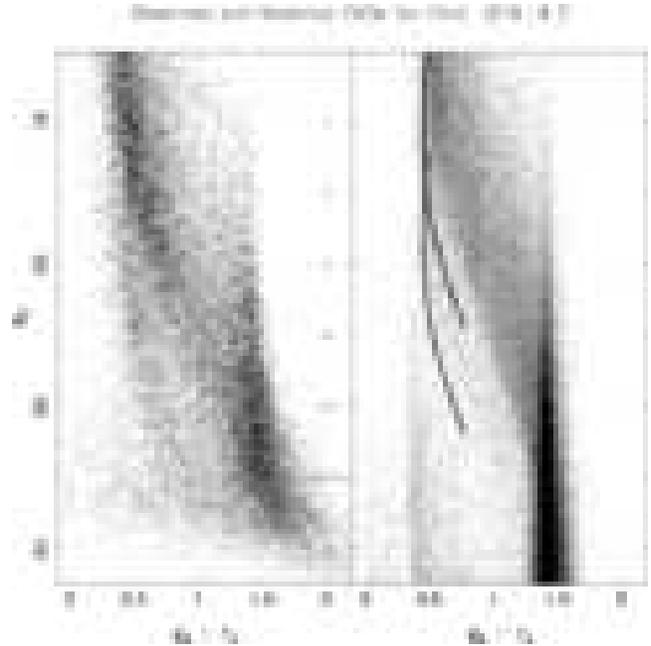}
\caption[]{\small Hess plot  of  ({\it l},{\it b})$\deg$ = (218,$+$6)$^\circ$ (Left) and the
  corresponding Besan\c{c}on model (Right). The figure is otherwise the same
  as Figure~\ref{fig195m20}. The sequences fitted here are offset
  by $-$1.5 mag and 0.0$^{+0.35}_{-0.3}$ mag. The heliocentric distance of these
  offsets are $\sim$5.5 \kpc\ and 11.0$^{+1.9}_{-1.4}$ \kpc. The CMD is 50\%
  complete at $g_\circ$ =  23.4 mag.
\label{fig218p6}}
\end{figure}

At the same Galactic longitude as the original Monoceros Ring detection made by
\citet{2002ApJ...569..245N}, the CMD is presented in Figure~\ref{fig218p6}. The two fiducials are offset by $-$1.5
mag for the brighter sequence and 0.0$^{+0.35}_{-0.3}$ mag for the
fainter coherent sequence. These result in distance estimates of $\sim$5.5\kpc\ and
$\sim$11.0$^{+1.9}_{-1.4}$ \kpc\ respectively. The CMD is 50\%
  complete at $g_\circ$ =  23.4 mag. Given the broad nature of the
Milky Way sequence, only the distance to the lower boundary is
stated. The S/N estimate for the MRi in this field is $\sim$34. This value
is higher than expected, probably due to the close proximity of the
dominant Main Sequence. The model does not cleanly subtract this
feature and so some counts remain to boost the signal to noise estimate. 

\subsubsection{Fields at \bf$(220,-15)^\circ$}\label{220m15des}
\begin{figure}\centering
\includegraphics[width=\hsize,angle=0]{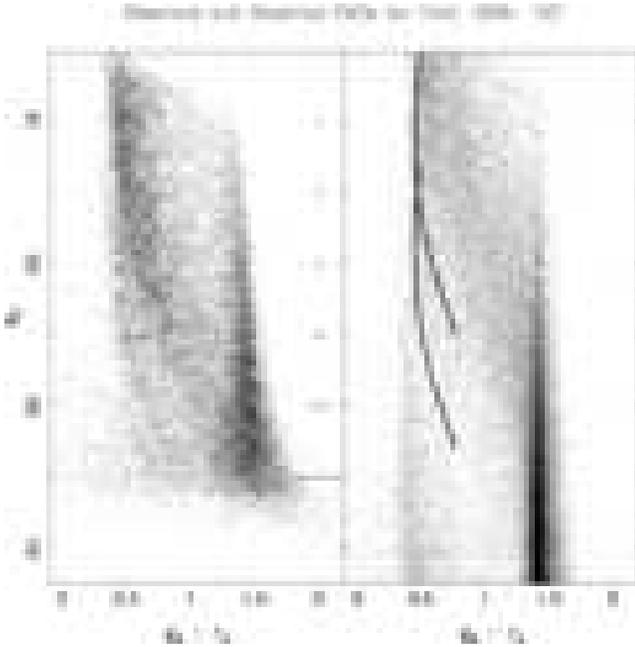}
\caption[]{\small Hess plot  of  ({\it l},{\it b})$\deg$ = (220,$-$15)$^\circ$
  (Left) and the corresponding Besan\c{c}on model (Right). The figure is otherwise the same
  as Figure~\ref{fig195m20}. The sequences fitted here are offset
  by $-$1.4 and 0.2$^{+0.3}_{-0.3}$ mag. The heliocentric distance of these
  offsets are 5.8 and 12.1$^{+1.7}_{-1.6}$ \kpc. At $g_\circ$ =
  22.9 magnitudes, the CMD is 50\% complete. 
\label{fig220m15}}
\end{figure}

As the Southern counterpart for the original Monoceros Ring detection made by
\citet{2002ApJ...569..245N}, the CMD for (220,$-$15)$\deg$ is
presented in Figure~\ref{fig220m15}. The two fiducials are offset by $-$1.4
mag for the brighter stronger sequence and 0.2$^{+0.3}_{-0.3}$ mag for
the fainter coherent sequence. These result in distance estimates of $\sim$5.8\kpc\ and
$\sim$12.1$^{+1.7}_{-1.6}$ \kpc\ respectively. At $g_\circ$ =
22.9 magnitudes, the CMD is 50\% complete. The fainter sequence is
found to have a S/N $\sim$ 32, although this estimate is probably contaminated by
the nearby brighter Main Sequence.

\subsubsection{Fields at \bf$(220,+15)^\circ$}\label{220p15des}
\begin{figure}\centering
\includegraphics[width=\hsize,angle=0]{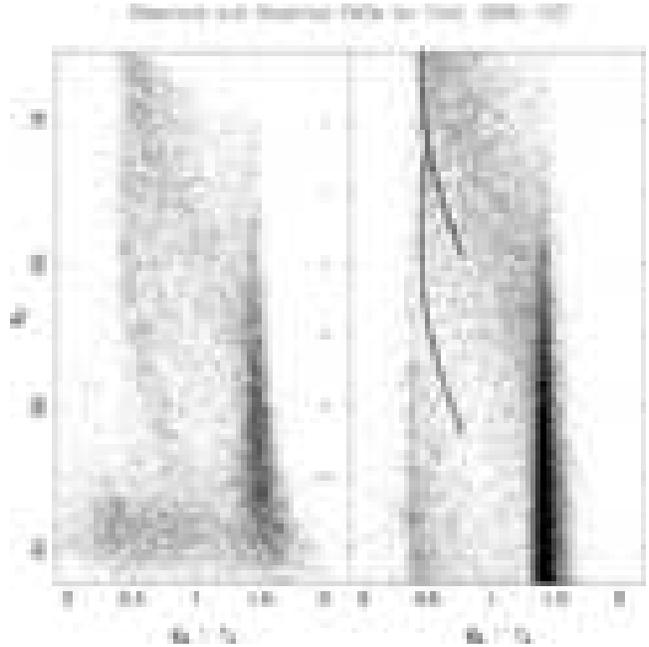}
\caption[]{\small Hess plot  of  ({\it l},{\it b})$\deg$ = (220,$+$15)$^\circ$ (Left) and the
  corresponding Besan\c{c}on model (Right). The figure is otherwise the same
  as Figure~\ref{fig195m20}. The sequences fitted here are offset
  by $-$2.5 and 0.0$^{+0.3}_{-0.35}$ mag. The heliocentric distance of these
  offsets are 3.5 and 11.0$^{+1.6}_{-1.6}$ \kpc. The 50\% completeness in
$g_\circ$ for this CMD is 23.6 mag.
\label{fig220p15}}
\end{figure}

This field, Figure~\ref{fig220p15}, is the closest to the original Monoceros Ring detection made by
\citet{2002ApJ...569..245N} taken during the present
survey. The two fiducials are offset by $-$2.5
mag for the brighter stronger sequence and 0.0$^{+0.3}_{-0.35}$ mag
for the fainter sequence. These result in distance estimates of $\sim$3.5\kpc\ and
$\sim$11.0$^{+1.6}_{-1.6}$ \kpc\ respectively. Being only five degrees from the fields presented
in \citet{2002ApJ...569..245N}, the distance to the MRi here is the
same as their distance estimate of 11 \kpc. The 50\% completeness in
$g_\circ$ for this CMD is 23.6 mag. A S/N $\sim$ 14 is found for the
fainter sequence.

\subsection{Canis Major Region}\label{CMaregion}
The following fields are part of a strip of observations linking
$(240,-9)^\circ$ to $(240,+10)^\circ$. The fields from {\it b} =
$-$6$\deg$ to {\it b} = $+$4$\deg$ are single pointings, while $(240,+6)^\circ$ has four pointings filling
out a $\sim$1$\deg \times$1$\deg$ field and $(240,+10)^\circ$ has two
pointings. These observations provide a glimpse as to how the Galaxy
profile changes across the Galactic plane. The high dust extinction of the lower Galactic latitudes has distorted the CMDs;
however, there is still information in these fields and for this reason
they have been left in.  The dust is more prominent in the Southern
fields as characterised by the lesser distortion of the CMDs in the Northern fields. All of
the fields from {\it b} = $-$6$\deg$ to {\it b} = $+$6$\deg$ were observed under the same conditions
with the $(240,+10)^\circ$ field observed the following night.

\subsubsection{Fields at \bf$(240,-9)\deg$}\label{240m9des}
\begin{figure}\centering
\includegraphics[width=\hsize,angle=0]{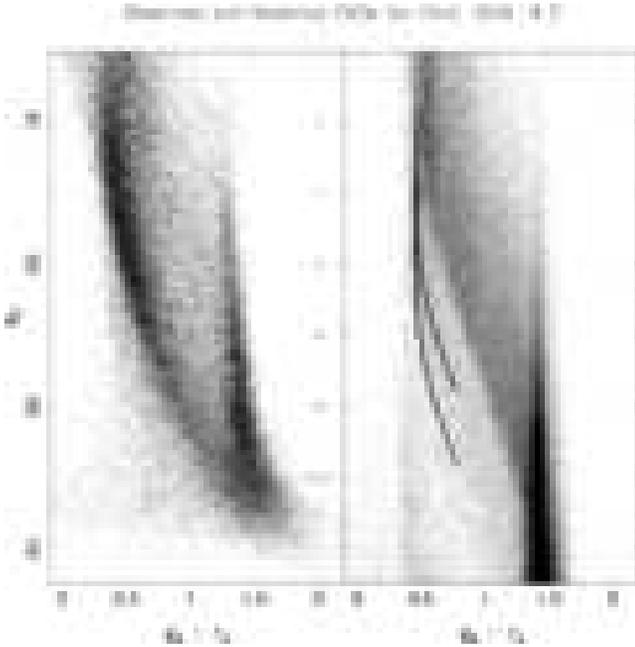}
\caption[]{\small Hess plot  of ({\it l},{\it b}) = (240,$-$9)$^\circ$ (Left) and the
  corresponding Besan\c{c}on model (Right).  As for
  Figure~\ref{fig195m20}. The strong sequence in this field is
  determined to be at $\sim$8.3 \kpc\ from an offset of $-$0.6
  mag with the fiducial sequence. Below this strong sequence is an
  excess of stars at a $g_\circ$ =  20-21 mag and ($g_\circ -
  r_\circ$) $\sim$ 0.5. This excess follows below the strong sequence
  as it increases in colour. Fitting a fiducial to this excess at an offset of
  0.4 mag a distance of 13.2 \kpc\ is obtained. This in excellent
  agreement with estimated distance to the MRi of 13.5 \kpc\ as
  derived in the AAT/2dF survey of \citet{2005MNRAS.364L..13C}. This
  is deemed a tentative detection of the MRi in this field. The CMD is 50\%
  complete at $g_\circ$ =  23.6 mag. Note the presence of BP stars at $g_\circ\lesssim18$.
\label{fig240m9}}
\end{figure}

This field is centred on the location of the putative core of the Canis Major
dwarf galaxy (Figure~\ref{fig240m9}). The very strong sequence running the length of the
CMD can be fit along the faint edge with a fiducial offset by
$-$0.6 mag which corresponds to $\sim$8.3 \kpc. Placing it roughly along
the centre of the feature requires an offset of $-$0.9 mag or $\sim$7.6
\kpc. In \citet{2005MNRAS.364L..13C}, the presence of the MRi in the
background of the CMa overdensity was revealed. This was determined to be at a distance of
13.5$\pm$0.3 \kpc. Indeed, just below the sequence dominating the
CMD there does seem to be an excess of stars which may be another coherent sequence, its contrast is low
due to the dominating effect of the CMa sequence. An offset of 0.4
mag is needed to align the fiducial with this feature,
corresponding to a heliocentric distance of 13.2 \kpc. Given the match
this makes with the AAT/2dF detection this is considered a tentative
detection. The CMD is 50\% complete at $g_\circ$ =  23.6 mag.

\subsubsection{Fields at {\bf$(240,-6)^\circ$}}\label{240m6des}
\begin{figure}\centering
\includegraphics[width=\hsize,angle=0]{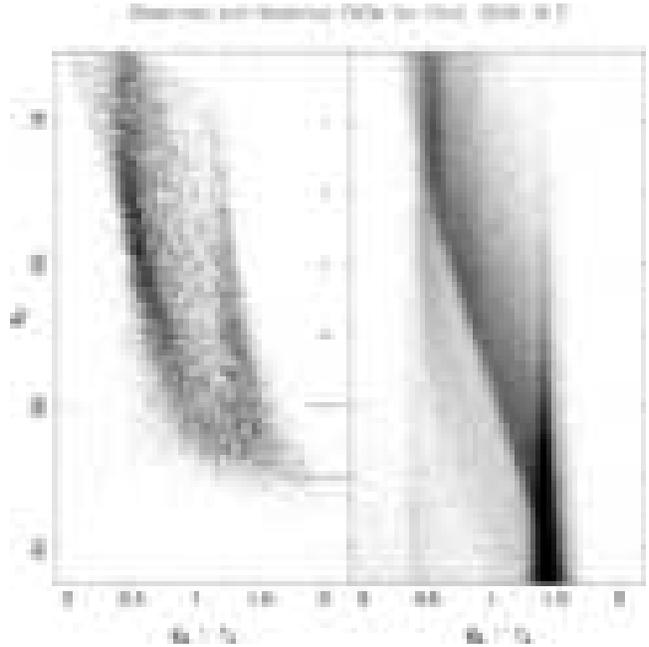}
\caption[]{\small As for Figure~\ref{fig195m20}, Hess plot of ({\it l},{\it
  b}) = (240,$-$6)$^\circ$ (Left) and the corresponding Besan\c{c}on
  model (Right). No offset is placed on this CMD due to the obvious
  distortion present which is most likely due to the increase in dust
  and non-photometric conditions. The CMD does show all the same
  features as present in the (240,$-$9)$^\circ$ field. Consisting of
  only one exposure, the 50\% completeness level has not been
  calculated. \label{fig240m6}}
\end{figure}

In Figure~\ref{fig240m6}, the strong sequence is still present in
this field although distorted by the increased level of dust and
possible non-photometric conditions. The Blue-Plume stars (see
$\S$~\ref{BPstars}) are still
easily visible and while there is contention over their origins, they
are still indicative that the general structures present here are
unchanged from the previous field. Consisting of only one exposure,
the 50\% completeness level has not been calculated. 

\subsubsection{Fields at \bf$(240,-4)^\circ$}\label{240m4des}
\begin{figure}\centering 
\includegraphics[width=\hsize,angle=0]{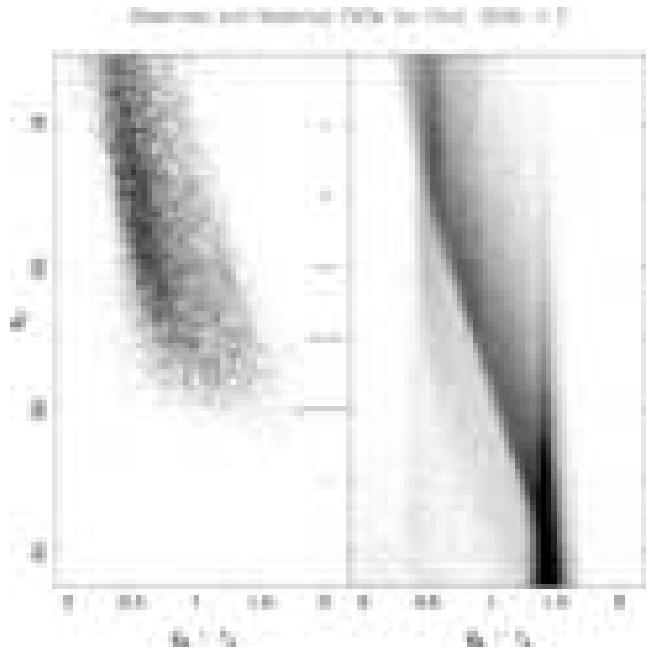}
\caption[]{\small Hess plots of  ({\it l},{\it b}) =
  (240,$-$4)$^\circ$ (Left) and the 
  corresponding Besan\c{c}on model (Right). The
  high dust levels in this field E(B-V)$\sim$1.4, has distorted the
  resultant CMD.  Although the exact features of the CMD are
  indistinguishable, the presence of the Blue Plume stars remains
  obvious, although seemingly weaker than the (240,$-$6)$\deg$ field. No
  completeness estimate has been made for this field.
\label{fig240m4}}
\end{figure}

This field (Figure~\ref{fig240m4}), as with those closest to plane, is heavily affected by
dust extinction. In particular, it has the second highest dust levels
in our survey, where E(B-V) is typically around 0.99. Again, despite the
loss of structure in this field, the CMD still shows evidence for a
Blue Plume population although it appears a little weaker than the
preceding (240,$-$6)$\deg$ field. The CMD has been left in the location
as determined by the calibration process of the CASU pipeline. The
high dust levels are the most likely cause  for the distortion on the
Main Sequences present in the CMD. No
completeness estimate has been made for this field.

\subsubsection{Fields at \bf$(240,-2)^\circ$}\label{240m2des}
\begin{figure}\centering
\includegraphics[width=\hsize,angle=0]{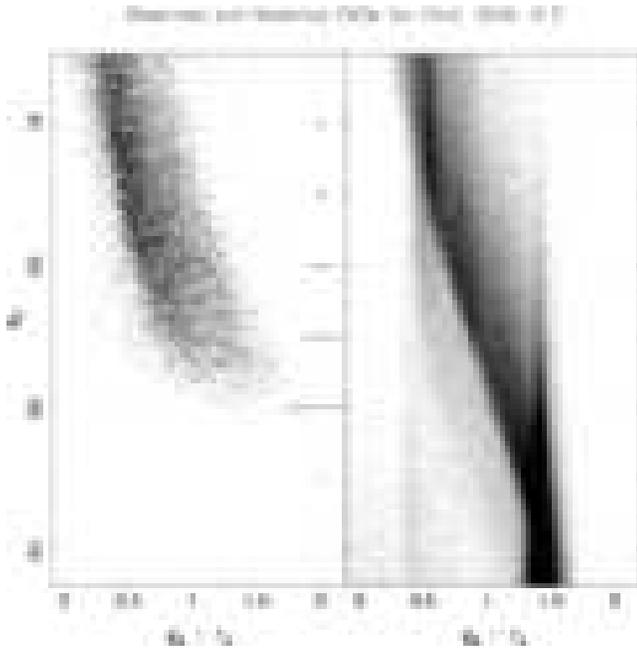}
\caption[]{\small Hess plots of  ({\it l},{\it b}) = (240,$-$2)$^\circ$ (Left)
  and its counterpart synthetic CMD (Right). The dust here again is heavily distorting
  the Main Sequences, allowing only the weak presence of the Blue Plume stars to
  be visible. As for the two previous fields, there is only one
  pointing in this direction and so no completeness estimate has been made.
\label{fig240m2}}
\end{figure}

The dust extinction in this field, Figure~\ref{fig240m2}, is extremely
high, typically around an E(B-V)$\sim$1.1. There is a faint suggestion of the
presence of Blue Plume stars, although less so than in the previous
field. The strong sequence visible in this region of sky is still apparent.
However, it is impossible to provide a distance estimate to this
structure. The limiting magnitude of this field is most likely heavily
affected by the dust accounting for its position with regard the CMDs
in the remaining survey locations. As for the two previous fields,
there is only one pointing in this direction and so no completeness
estimate has been made.

\subsubsection{Fields at \bf$(240,+2)^\circ$}\label{240p2des}
\begin{figure}\centering
\includegraphics[width=\hsize,angle=0]{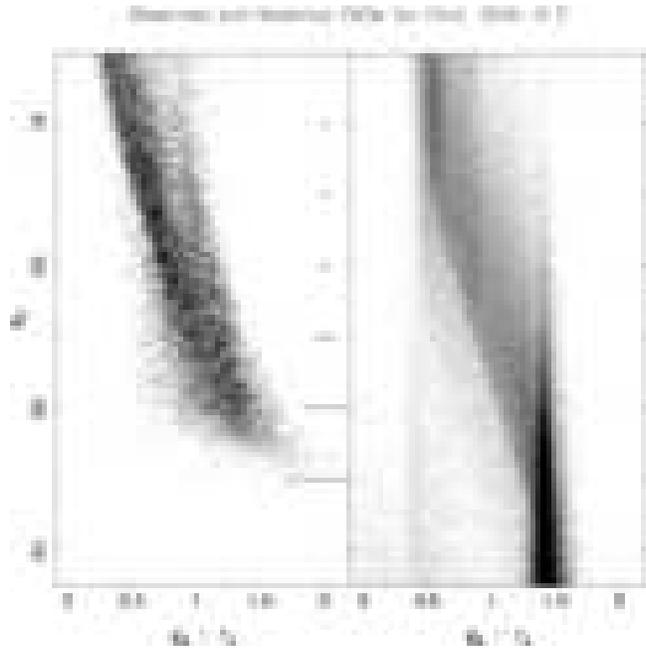}
\caption[]{\small Hess plots of  ({\it l},{\it b}) = (240,$+$2)$^\circ$ (Left)
  and its counterpart synthetic CMD (Right). Although heavily affected by dust,
  the dominant Main Sequence is now clearly visible. The Blue Plume population seems to
  have disappeared or become too bright, leaving what may be solely Galactic stars
  behind. No distance estimate has been attempted for this
  field. Despite the distortion, there is obviously a very strong sequence present in this field. Continuing the strip of single
  pointings above the plane, this field also has no completeness estimate.
\label{fig240p2}}
\end{figure}

This field, Figure~\ref{fig240p2}, begins to re-emerge from the dust problems below the plane,
showing a strong sequence across the CMD. Unfortunately, due to the distortion effects of the dust, an
estimate of the distance to this structure is still not possible. It is difficult to
judge whether the sequence feature seen here is created by the
same structure detected below the plane. The Blue Plume population is
no longer present, which may indicate these are purely Galactic
stars or that it has shifted to brighter magnitudes which are
saturated in this survey. Continuing the strip of single pointings
above the plane, this field also has no completeness estimate.

\subsubsection{Fields at \bf$(240,+4)^\circ$}\label{240p4des}
\begin{figure}\centering 
\includegraphics[width=\hsize,angle=0]{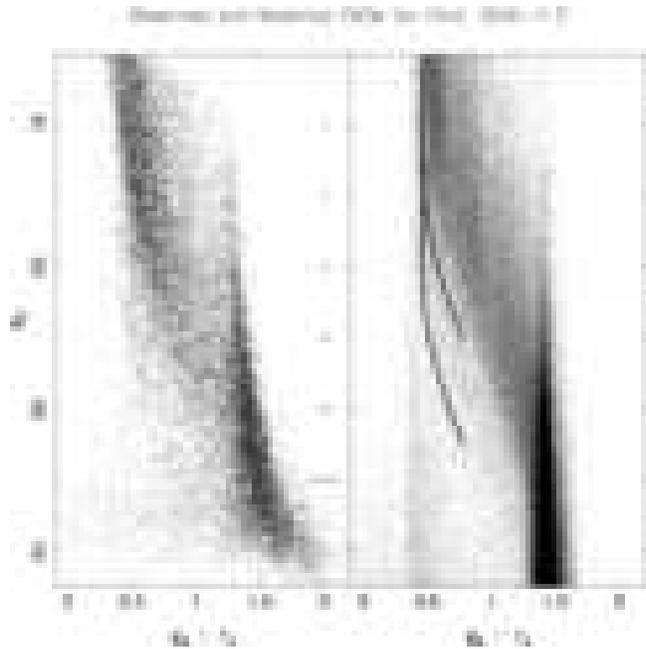}
\caption[]{\small Hess plots of  ({\it l},{\it b}) = (240,$+$4)$^\circ$ (Left). The
  fiducial sequences plotted on top of the Besan\c{c}on model
  (Right) are at offsets of $-$1.4 mag and 0.1 mag. These are at
  $\sim$5.8 \kpc\ and 11.5 \kpc, respectively. This field with a
  single pointing has no completeness estimate. 
\label{fig240p4}}
\end{figure}

Here the main components present in the CMD, Figure~\ref{fig240p4}, can be fit with fiducials
at magnitude offsets of $-$1.4 mag and 0.1 mag. The distances then to these
features are 5.8 \kpc\ and 11.5 \kpc, heliocentric. This field with a single pointing has no
completeness estimate.

\subsubsection{Fields at {\bf$(240,+6)^\circ$}}\label{240p6des}
\begin{figure}\centering
\includegraphics[width=\hsize,angle=0]{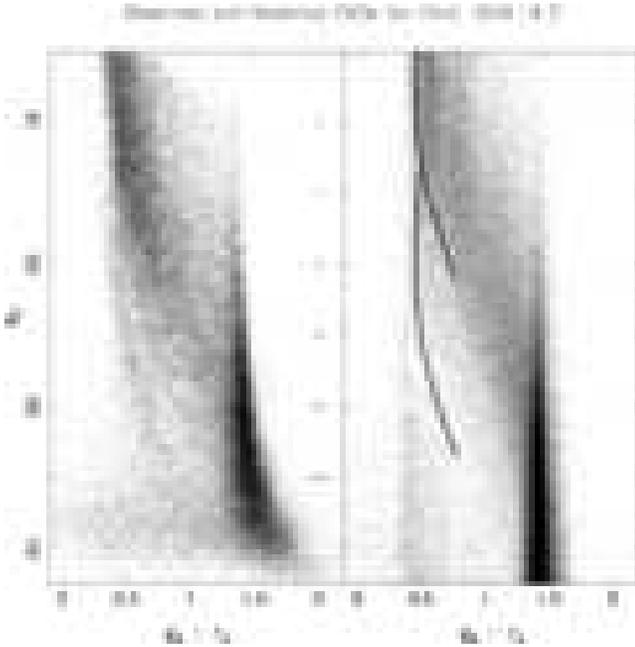}
\caption[]{\small Hess plots of  ({\it l},{\it b}) = (240,$+$6)$^\circ$ (Left)
  and its counterpart synthetic CMD (Right). As for Figure~\ref{fig195m20}. The two
  fiducials are placed at offsets of $-$1.2 mag and 0.3$^{+0.3}_{-0.3}$ mag, aligning
  with the features in the data. The heliocentric distance is 6.3
  \kpc\ and 12.6$^{+1.9}_{-1.6}$ \kpc, respectively. A 50\%
  completeness in this CMD is found at $g_\circ$ = 24.0.
\label{fig240p6}}
\end{figure}

Consisting of four pointings, this field is the largest of the Canis
Major region targets above the plane (Figure~\ref{fig240p6}).  The strong sequence here corresponds to a
magnitude offset of $-$1.2 mag, or $\sim$6.3 \kpc. With the lower sequence
residing at $\sim$12.6$^{+1.9}_{-1.6}$ \kpc\ or 0.3$^{+0.3}_{-0.3}$
mag of offset. A 50\% completness in this CMD is found at $g_\circ$ = 24.0.

\subsubsection{Fields at {\bf$(240,+10)^\circ$}}\label{240p10des}
\begin{figure}\centering
\includegraphics[width=\hsize,angle=0]{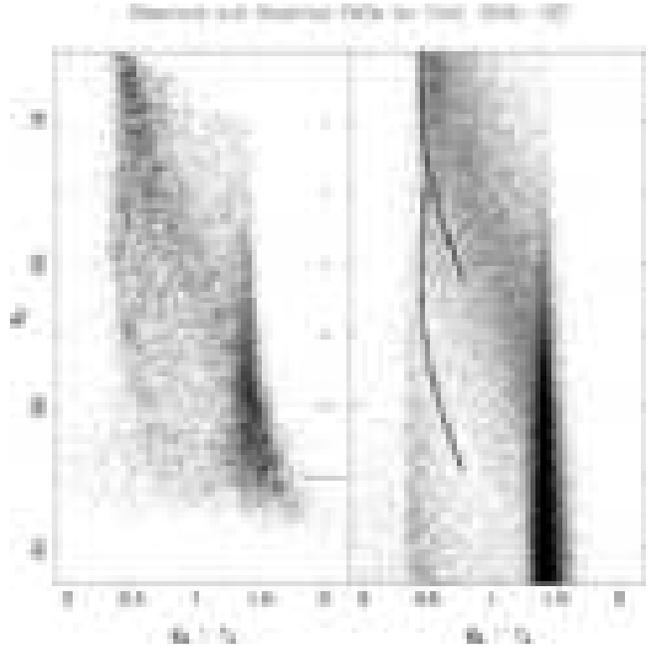}
\caption[]{\small Hess plots of  ({\it l},{\it b}) = (240,$+$10)$^\circ$
  (Left) and its counterpart synthetic CMD (Right). As for Figure~\ref{fig195m20}. There are
  two structures present in the data for which the fiducial sequences are placed onto the model. The brighter of the two is at a
  magnitude offset of $-$2.2 mag, or $\sim$4.0 \kpc, and the
  fainter fiducial with an offset of 0.5$^{+0.3}_{-0.3}$ mag, or 13.8$^{+2.1}_{-1.7}$
  \kpc. This field has a 50\% completeness in $g_\circ$ at 22.65. 
\label{fig240p10}}
\end{figure}

Two pointings make up this field (Figure~\ref{fig240p10}), and again two sequences are
clearly visible in the CMD. The closer of the two is at $-$2.2
mag of offset, or 4.0 \kpc. The more distant sequence is
offset at 0.5$^{+0.3}_{-0.3}$ mag, or 13.8$^{+2.1}_{-1.7}$ \kpc\
heliocentric. This field has a 50\% completeness in $g_\circ$ at
22.65. The S/N of the fainter feature is estimated to be $\sim$ 22.

\subsubsection{Fields at \bf$(245,-9)^\circ$}\label{245m9des}
\begin{figure}\centering
\includegraphics[width=\hsize,angle=0]{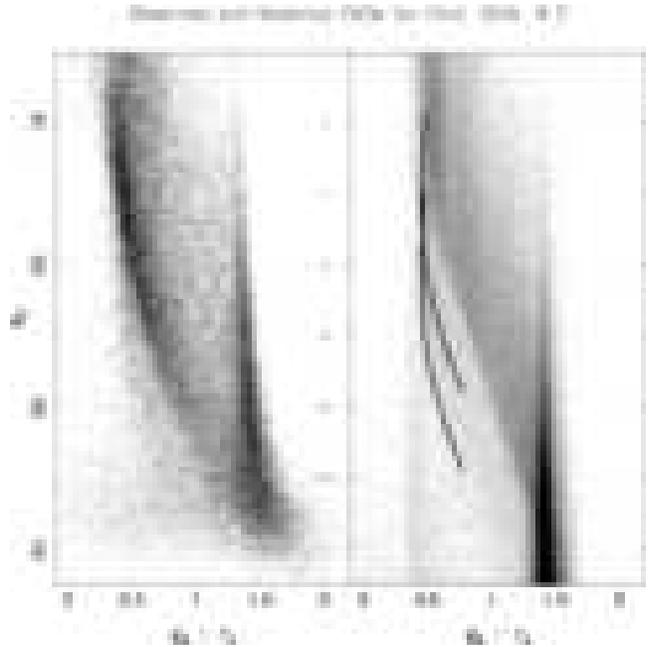}
\caption[]{\small Hess plot  of ({\it l},{\it b}) = (245,$-$9)$^\circ$ (Left) and the
  corresponding Besan\c{c}on model (Right). As for
  Figure~\ref{fig195m20}. As for the field (240,$-$9)$\deg$, this field
  contains a very strong sequence and is fit by the same offset
  of $-$0.6 mag or 8.3 \kpc. The presence of a fainter more distant sequence can be seen below the strong sequence and is fit with
  an offset of 0.5 mag at a heliocentric distance of 13.8 \kpc. At
  23.3 magnitudes in $g_\circ$, the CMD is 50\% complete.  Note the presence of BP stars at $g_\circ\lesssim18$. \label{fig245m9}}
\end{figure}

This field, presented in Figure~\ref{fig245m9}, was chosen to probe the lateral extent of the Canis Major
dwarf/overdensity. Being 5$\deg$ East in Longitude, the dominant sequence in this region of sky is still very strong.  The
fiducial sequence has been fitted with an offset of $-$0.6 mag. This offset
corresponds to a distance of $\sim$8.3 \kpc. As with the field
(240,$-$9)$^\circ$, a fiducial closer to the centre of this feature is
offset by $-$0.9 mag or 7.3 \kpc. This value matches the distance to
dwarf galaxy as reported by \citet{2004MNRAS.355L..33M}. A fiducial has been fitted to the faint MRi population just below the strong
sequence. The offset of 0.5 mag corresponds to 13.8 \kpc\
heliocentrically. At 23.3 magnitudes in $g_\circ$, the CMD is
50\% complete.

\subsection{Maximum Warp region}
The remaining fields in this part of the survey occupy the part of the
Galaxy which is most affected by the Galactic warp. Due to
predominantly poor weather, only three fields of this regions are
presented. However, they do contain information regarding the Monoceros
Ring and the Canis Major dwarf and also continue the Galactic plane
survey into the fourth quadrant.
\subsubsection{Fields at \bf$(260,-10)\deg$}\label{260m10des}
\begin{figure}\centering
\includegraphics[width=\hsize,angle=0]{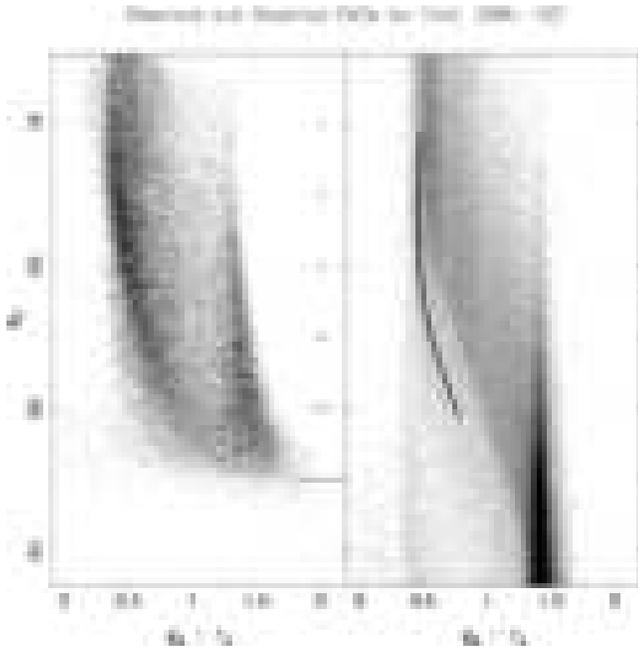}
\caption[]{\small Hess plot  of ({\it l},{\it b}) = (260,$-$10)$^\circ$ (Left) and the
  corresponding Besan\c{c}on model (Right). As for
  Figure~\ref{fig195m20}. As for the fields nearer the (240,$-$9)$\deg$ field, this field
  contains a very strong sequence and is fit by the an offset
  of $-$0.3 mag or 9.6 \kpc. The MRi can not be seen in this field. The 50\%
completeness level in $g_\circ$ is found at 22.8 magnitudes for this field.
\label{fig260m10}}
\end{figure}

This field is located 20$\deg$ from the Canis Major
dwarf/overdensity and although the Blue Plume stars have maintained
their location on the CMD, the strong sequence has shifted to
fainter magnitudes (Figure~\ref{fig260m10}).  The fiducial sequence
has been fitted with an offset of $-$0.3 mag. This offset
corresponds to a distance of $\sim$9.6 \kpc. There is no clear
evidence of the MRi in this field, which would reside below the strong
sequence close to the limiting magnitude of this data. The 50\%
completeness level in $g_\circ$ is found at 22.8 magnitudes for this
field. \citet{2006MNRAS.367L..69M} report on the detection of the MRi
12 degrees further in latitude from this field, in front of the Carina
dwarf galaxy. This detection was made in velocity space and currently
there is no firm estimate of the distance to the MRi at this
longitude. Until observations in this region are extended to fainter
magnitudes no detection of the MRi can be reported in this field.

\subsubsection{Fields at \bf$(273,-9)^\circ$}\label{273m9des}
\begin{figure}\centering
\includegraphics[width=\hsize,angle=0]{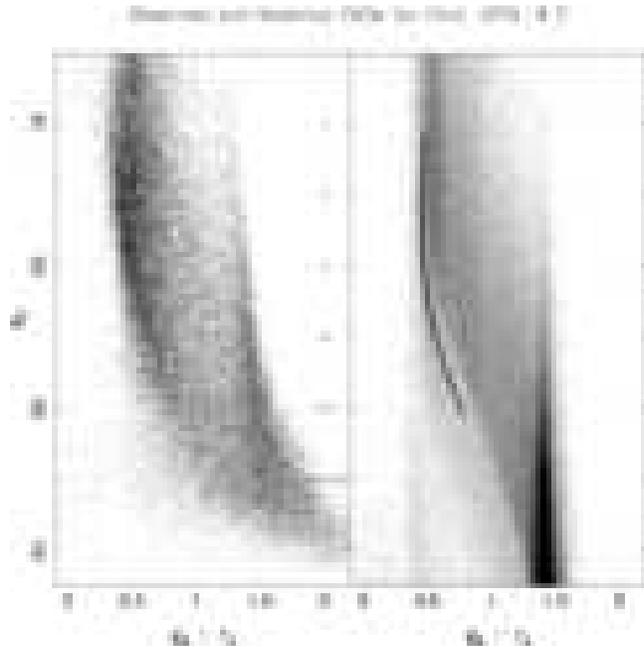}
\caption[]{\small Hess plot  of ({\it l},{\it b}) = (273,$-$9)$^\circ$ (Left) and the
  corresponding Besan\c{c}on model (Right). As for
  Figure~\ref{fig195m20}. As for the field (260,$-$10)$\deg$, this field
  contains a very strong sequence and is fit by a similar offset
  of $-$0.3 mag or $\sim$9.6 \kpc. The CMD is 50\% complete at $g_\circ$ =  23.4 mag. The distortion of
  this field due to dust means the fiducial cannot be fitted
  accurately. 

\label{fig273m9}}
\end{figure}

This field is at the same latitude as the Canis Major dwarf/overdensity and,
as seen in the previous field, the strong sequence in this field
has shifted to fainter magnitudes (Figure~\ref{fig273m9}). The
dominant sequence remains very strong and is fitted with a fiducial with an
offset of $-$0.3 mag. This offset corresponds to a distance of 
$\sim$9.6 \kpc. The CMD is 50\% complete at $g_\circ$ =  23.4 mag. This field shows some distortion at its faintest
extremes, most likely due to a combination of the levels of dust in
the region and/or changing conditions during the observations. While
the fiducial has been fitted to the data, it is clear that the
distortion inhibits the ability to accurately locate the distance. Of
most importance is the presence of the strong sequence in this
field and the Blue Plume stars which reside in the same location as
the fields at different longitudes.

\subsubsection{Fields at \bf$(276,+12)^\circ$}\label{276p12des}
\begin{figure}\centering
\includegraphics[width=\hsize,angle=0]{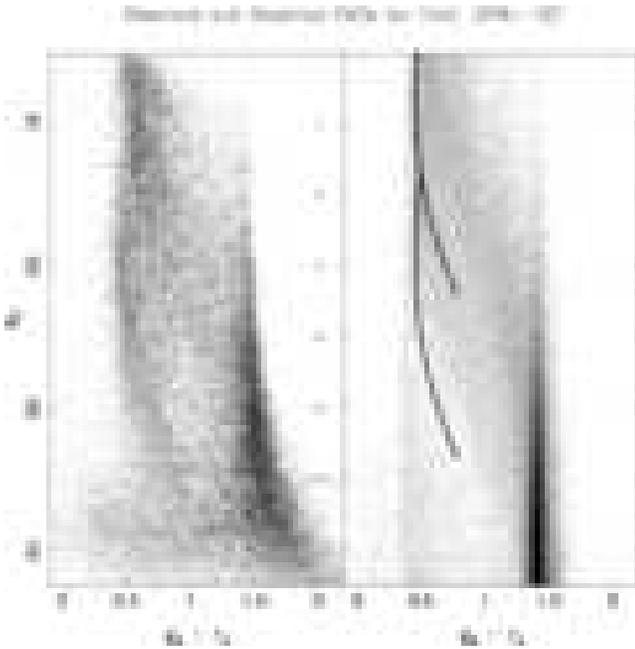}
\caption[]{\small Hess plot  of ({\it l},{\it b}) = (276,$+$12)$^\circ$ (Left) and the
  corresponding Besan\c{c}on model (Right). As for
  Figure~\ref{fig195m20}. As for the field (260,$-$10)$\deg$, this field
  contains a very strong sequence and is fit by a similar offset
  of 0.3 mag or $\sim$12.6 \kpc. The Milky Way/Dominant 
  sequence is found at 4.4 \kpc\ through an offset of $-$2.0
  mag. This CMD is 50\% complete at $g_\circ$ =  23.7 mag.
\label{fig276p12}}
\end{figure}

In Figure~\ref{fig276p12}, the field shows clear
evidence for the MRi.  The fiducial sequence
has been fitted to the MRi with an offset of 0.3 mag. This offset
corresponds to a distance of $\sim$12.6 \kpc. The Milky Way/Dominant sequence is found with a fiducials corresponding to $-$2.0 mag or 4.4 \kpc\ heliocentric. This CMD is 50\%
  complete at $g_\circ$ =  23.7 mag. The fainter sequence is estimated
  to have S/N $\sim$ 26.

\section{Analysis \& Discussion}\label{discussionaat}
This section discusses the detections of the Monoceros Ring and
those in the Canis Major region. This is done separately as there
is no consensus on either the origin of these structures or their
connectedness. 
\begin{figure}\centering
\includegraphics[width=\hsize,angle=270]{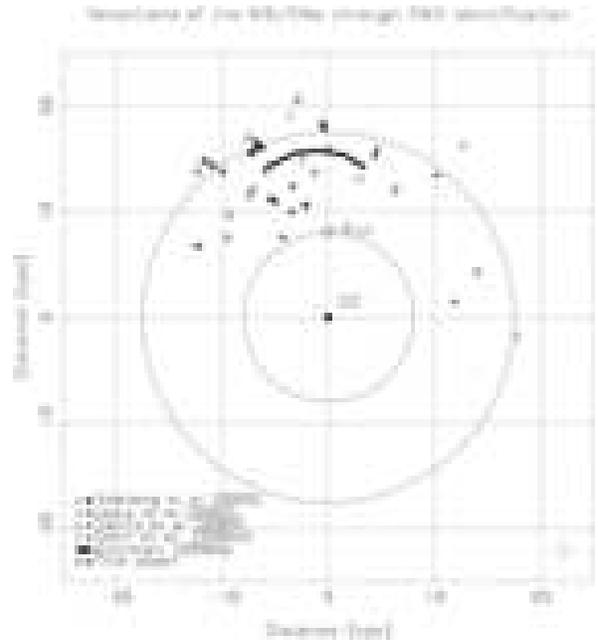}
\caption[]{\small Location of all of the detections of
  the Monoceros Ring (to date) and Canis Major dwarf (from this paper)
  projected onto the Galactic plane. The symbols denote the source of
  the distances and locations presented here. Filled symbols are
  detections made above the plane of the Galaxy, with empty symbols
  for those fields below the Galactic plane. Approximate locations of
  the MRi stream (above the plane) seen in \citet{2006ApJ...651L..29G} are illustrated with the solid line. The
  stars (filled and empty) are those detections outlined in this
  survey. A $\pm$1 \kpc\ error bar in the lower right corner is
  indicative of the accuracy of these detections. The strip of observations at {\it l} = 240$\deg$ is clearly
  evident with the fields below the plane (open stars) at the distance
  of the Canis Major overdensity and the fields above the plane
  (filled stars) at the Monoceros Ring distance.  Since each field in
  the survey has been fit with two fiducial sequences, each is present
  here to illustrate their location with regard the other detections.
\label{figalldetect}}
\end{figure}

\subsection{The Monoceros Ring}
From this survey, we interpret the presence of apparent coherent sequences
beyond the edge of the Thin/Thick disc as the Monoceros Ring. This results in 10 detections across the entire
longitude range of this survey and on both sides of the
Galactic plane. Importantly, these results show that the Monoceros
Ring cannot be reproduced by a warped disc
scenario leaving only the Galactic flare or a non-Galactic origin for these
stars. Figure~\ref{figalldetect} shows all the previous MRi detections
as projected onto the Galactic plane.  The star symbols are the
detections outlined in Table~\ref{ResultsTable} and so those star
symbols close to the Sun are MW/CMa sequence distances, while
those further away are the MRi. Fields above the plane are filled
symbols and those below the plane are empty symbols.

The circles at (61,$\pm$15)$\deg$ and (75,$+$15)$\deg$
from \citet{2005MNRAS.362..475C} are included here as detections. A
review of these fields during the preparation of this paper has found
them to be detections of the MRi. The previous use of the Besan\c{c}on
synthetic Galaxy model out to only 50 \kpc\ introduced a turn-off in
the Halo population near these MRi sequences. This added confusion to the identification of the
MRi in these fields but increasing the model distance limits to 100 \kpc\ removed this
turn-off and showed these features to be truly MRi detections.

The previous detections of the Monoceros Ring from CMDs, seen in
Figure~\ref{figalldetect}, show the distance estimates do not form a neat coherent picture of the
structure. Rather, there are many gaps between detections which limit
our understanding of the overall shape of this feature. Our results
seem to show a slight systematic preference for fields below the plane
being further away than those above the plane. The fields
above the plane from about {\it l} = (118 - 240)$\deg$ do seem to
maintain a consistent galactocentric distance of about 16 - 17 \kpc. There
are also a few detections below the plane which consistently fall
outside of the 16 \kpc\ circle. 

The difficulty with Figure~\ref{figalldetect} is that while it
distinguishes between those fields above and below the plane, it does
not fully convey how far apart those fields truly are. For instance,
in the {\it l} = 220$\deg$ direction, the original detection by
\citet{2002ApJ...569..245N}, the (218,$+$6)$\deg$ field and the
(220,$+$15)$\deg$ reside at the same distance. The (220,$-$15)$\deg$
field with $\sim$30$\deg$ of latitude between it and the
\citet{2002ApJ...569..245N} detection is found slightly further away
at $\sim$12.1 \kpc. This is same structure present on both
sides of the plane and with no young stars forming part of the MRi it
is unlikely to have been caused by perturbations within the Disc.

\subsubsection{Can the Flare explain the Ring?}\label{flare}
\begin{figure}\centering
\includegraphics[width=\hsize,angle=0]{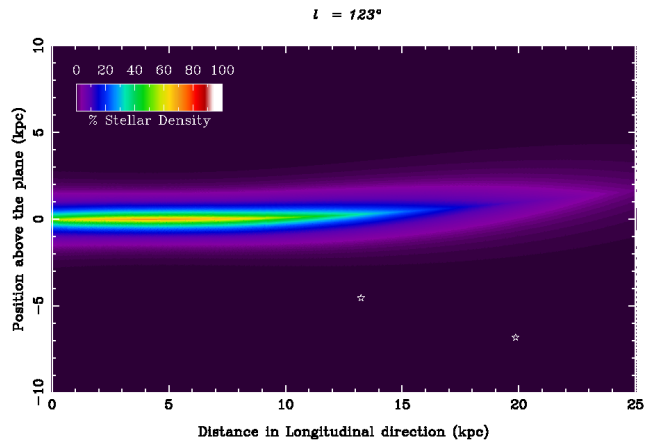}
\caption[]{\small Stellar density profile of the Galaxy as described
  by \citet{Yusi2004}, in the direction of {\it l} = 123$\deg$. It
  covers all Galactic latitudes from $-$90$\deg<${\it }b$<$90$\deg$. The star
  symbols denote the location of the Monoceros Ring detections as
  presented in \citet{2005MNRAS.362..475C}. The colour scale
  corresponds to the stellar density of a given coordinate divided by the
  maximum density for the region. The Galactic plane is seen clearly
  as the region of high density and it shifts above the plane with
  increasing distance in correspondence with the properties of the Northern warp.
\label{fig123slice}}
\end{figure}

\begin{figure}\centering
\includegraphics[width=7cm,angle=270]{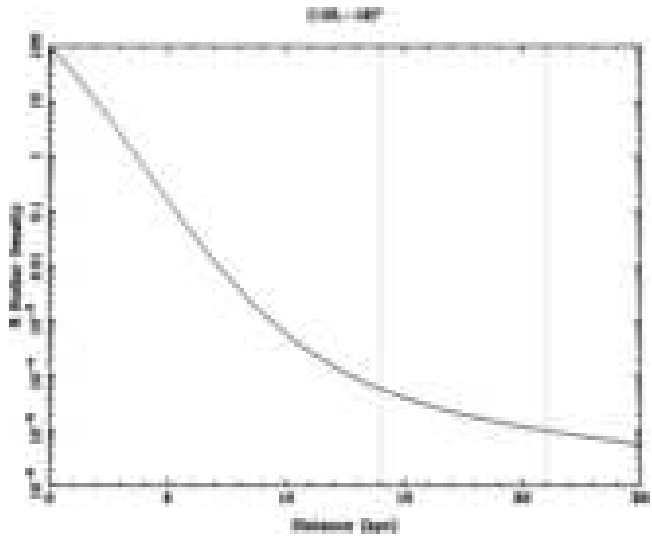}
\caption[]{\small Line-of-sight stellar density profile
  from the \citet{Yusi2004} model for (123,$-$19)$\deg$. The location of the two vertical lines correspond
  to the distances of the two MRi detections in this direction. The
  first is located at 14 \kpc\ with the second at 21 \kpc. This shows
  that the expected stellar density contribution due to the Warp and
  Flare is 10$^{-4}$ times smaller than the maximum stellar density along
  that line-of sight. There is also no increase in density to account for the presence
  of the MRi at these distances in this direction.
\label{fig123point}}
\end{figure}

\citet{2006A&A...451..515M} present a possible explanation for the existence of the
Monoceros Ring in terms of the flaring of the Galactic disc. The Flare
in the Disc is in addition to the Warp and represents a gradual
widening of the Disc with increasing Galactic radii.  They present several
figures showing the location of the previous MRi detections, from both
the \citet{2002ApJ...569..245N} and \citet{2005MNRAS.362..475C}, as
intersecting the Flare in the \citet{Yusi2004} model. Figure~\ref{fig123slice} shows the
Yusifov models' prediction for the stellar density profile of the Galaxy in
direction of {\it l} = 123$\deg$, a field which contains two
detections of the MRi in the same direction, namely {\it b} =
$-$19$\deg$. The Warp in this part of the Galaxy extends North as seen
here and because of the Flare, the width of the Galaxy grows with
increasing distance. The Sun is located at the origin. The stellar
density can be seen to decrease in both the radial and
the perpendicular directions. The distance estimates to the detections
found in this field, from \citet{2005MNRAS.362..475C}, are shown as
stars. There is an error of at least 10\% on their location, as
explained in that paper, and in Figure~\ref{fig123slice}, this error aligns itself with the line
joining the location of the symbol and the origin. This figure clearly
shows that density in the region of the detections is below 0.5\% of the
maximum stellar density in this direction. Attributing this feature
to the Warp is not feasible since a warp would only produce a single
smooth Main Sequence. The Flare also does not introduce any structures or boundaries
of significance in the entire region and does not seem suitable for
explaining the MRi in this direction. To investigate this further,
Figure~\ref{fig123point} shows the line of sight density profile for
the (123,$-$19)$\deg$ field from the model. The two vertical lines are
placed at the distances of the two detections. It is evident
that the model does not show any increase in density with increasing
distance from the Sun and that the Flare cannot explain the presence
of the Monoceros Ring here.    

\begin{figure}\centering
\includegraphics[width=\hsize,angle=0]{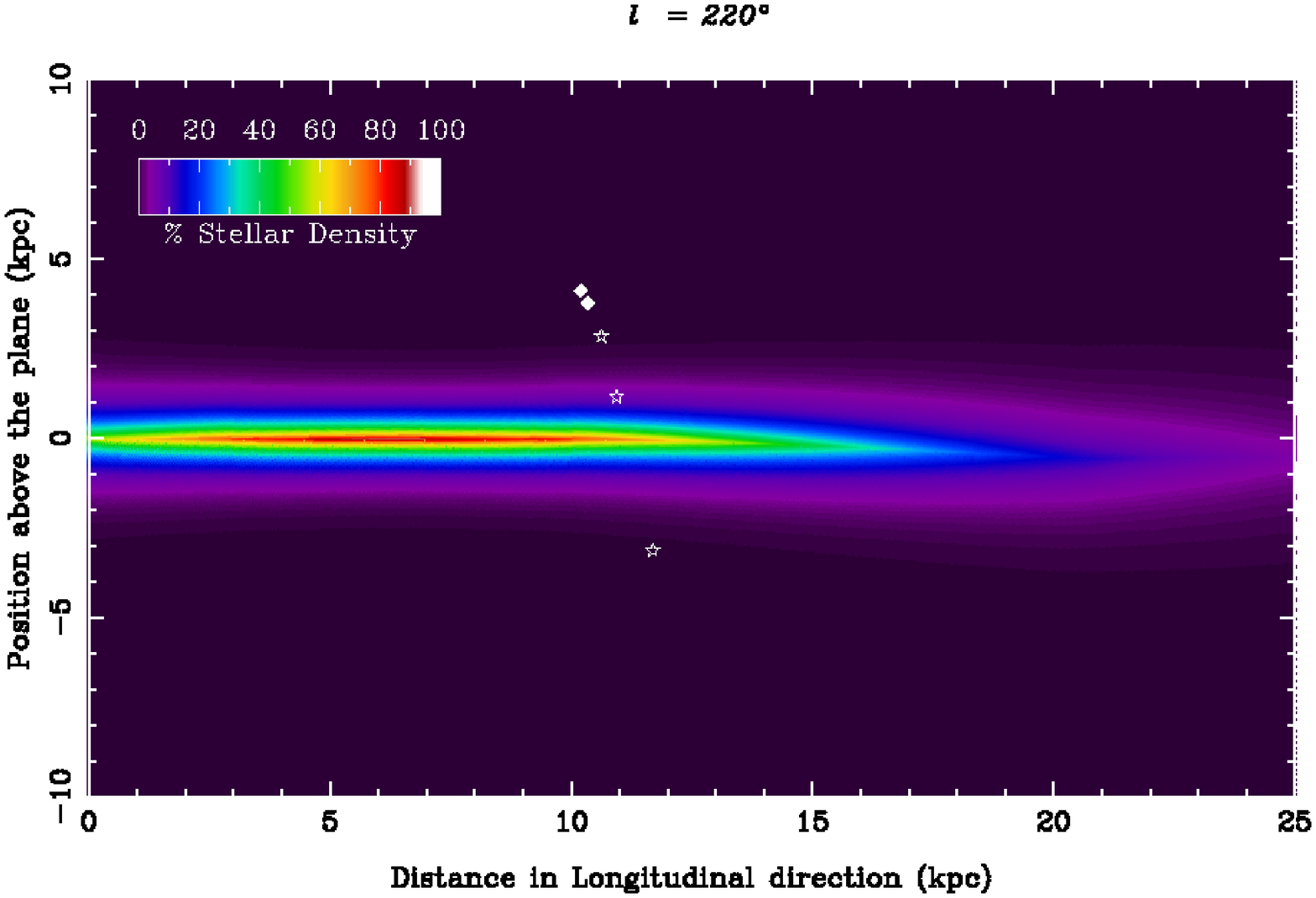}
\caption[]{\small As per Figure~\ref{fig123slice}, this is the
  stellar density profile of the Galaxy in the direction {\it l} =
  220$\deg$. The star symbols corresponds the detections of the MRi as
  outlined in this paper, while the diamonds are from \citet{2002ApJ...569..245N}.
\label{fig220slice}}
\end{figure}

\begin{figure}\centering
\includegraphics[width=\hsize,angle=0]{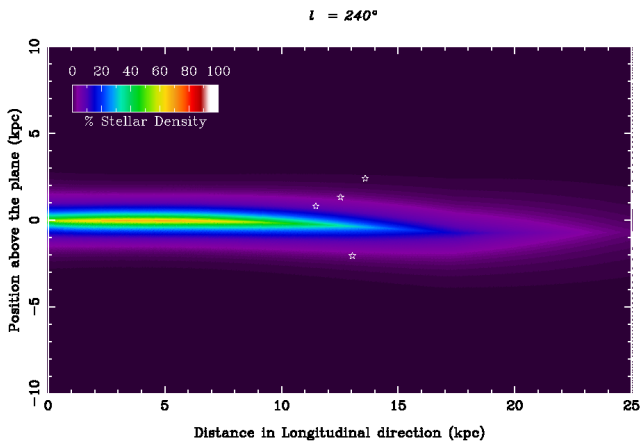}
\caption[]{\small As per Figure~\ref{fig123slice}, this is the
  stellar density profile of the Galaxy in the direction {\it l} =
  240$\deg$. The star symbols corresponds the detections of the MRi as
  outlined in this paper. 
\label{fig240slice}}
\end{figure}

\begin{figure}\centering
\includegraphics[width=7cm,angle=270]{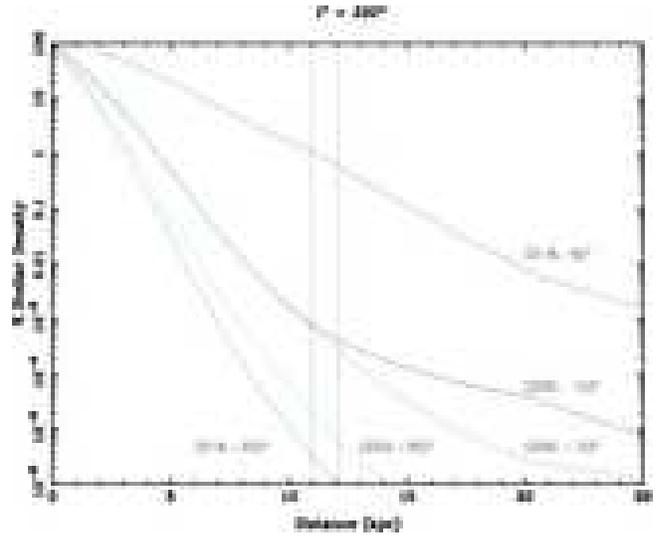}
\caption[]{\small Line-of-sight stellar densities for each
  detection around {\it l} = 220$\deg$. The two vertical line delimit
  the maximum and minimum distance estimates for the MRi in this
  region. As with Figure~\ref{fig123point}, the overdensity of stars
  belonging to the MRi do not originate from the warp or flare of the Galaxy.
\label{fig220line}}
\end{figure}

Two other regions of interest when interpreting the Monoceros Ring are at
{\it l} = 220$\deg$ and {\it l} = 240$\deg$.  In both cases, the MRi has
been detected above and below the plane at the same longitude. The
stellar density predictions from the Yusifov model are shown in
Figures~\ref{fig220slice} \& \ref{fig240slice}. In
Figure~\ref{fig220slice}, the diamond symbols are from
\citet{2002ApJ...569..245N}, while the star symbols are from this
paper and altogether they show that the MRi in this region is an extended
vertical structure in the disc.  Although it is unsure whether the
detections on either side of the plane are of the same origin, the
coincidence in their distances should be noted. It is also apparent
that known Galactic structure is unable to account for an overdensity
at this distance. Could the density profile seen in
Figure~\ref{fig123point} be peculiar to that direction?
Figure~\ref{fig220line} presents the line of sight density for all the
detections seen in Figure~\ref{fig220slice}. The two vertical lines
delimit the minimum and maximum MRi distance estimates for these
fields. Again, it can be seen that the stellar density profile is
unable to reproduce an increase in stellar density which could explain
the MRi.  The MRi detections here are found within a variety of
stellar densities and so neither the Warp or the Flare can be invoked
to justify their presence. At {\it l} = 240$\deg$, the Northern
detections range from $\sim$11.5 \kpc\ at (240,$+$4)$\deg$ to $\sim$13.8
\kpc\ at (240,$+$10)$\deg$. Below the plane, the field (240,$-$9)$\deg$ is
also found at $\sim$13.8 \kpc\ which is consistent with the results of
\citet{2005MNRAS.364L..13C}. While there is not the neat correlation
between the Northern and Southern detections in this region, as with
the {\it l} = 220$\deg$ fields, it does confirm again that the smooth
stellar density distributions are unable to account for the
overdensity of stars contained in the MRi. The warped flared Milky Way
does not contribute significantly at the locations of the MRi, as
claimed by \citet{2006A&A...451..515M}.

There is no neat coherent picture of the MRi structure even considering
the many detections of it, primarily due to relatively small areas of sky surveyed, combined
with rough distance estimates. However, regardless of this,
the Galactic flare is not a likely source for this overdensity of
stars. And while this has only been tested on the Yusifov model, the
similarities between the various models make it unlikely that any of
them will be able to generate a significant overdensity of stars at
the distances found in the data. Additionally, even the possibility of
it being a distant spiral arm, is countered by the vertical extent of
the MRi out of the plane. For those dissatisfied with a tidal stream
origin for the MRi, other mechanisms will have to be invoked. In the
meantime, the best explanation is a tidal stream scenario and so given
that, where then is the progenitor?



\subsection{The Canis Major Dwarf}\label{dwarf}
\begin{figure}\centering
\includegraphics[width=7cm,angle=270]{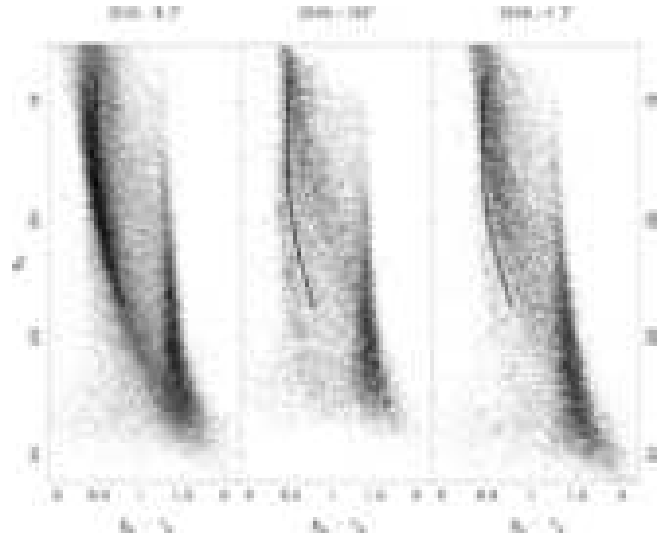}
\caption[]{\small Colour-Magnitude Diagrams of
  (240,$-$9)$\deg$ (Left), (240,$+$10)$\deg$ (Centre) and (240,$+$4)$\deg$
  (Right). The fiducial sequence placed on each figure is for the
  distance of 7.2 \kpc. If there was no warp in the disc of the Milky
  Way then the (240,$-$9)$\deg$ and (240,$+$10)$\deg$ should be
  similar due the approximate symmetry of the disc. However, 
  if the Milky Way is warped by $\sim$3$\deg$ as suggested by
 \citet{2006A&A...451..515M} then (240,$-$9)$\deg$ and (240,$+$4)$\deg$ should be
  similar. Given that the two fields will differ with increasing
  heliocentric radius, the fiducial is placed at the same distance to
  highlight where they should be the same. 
\label{figthreehess}}
\end{figure}
\begin{figure}\centering
\includegraphics[width=7cm,angle=270]{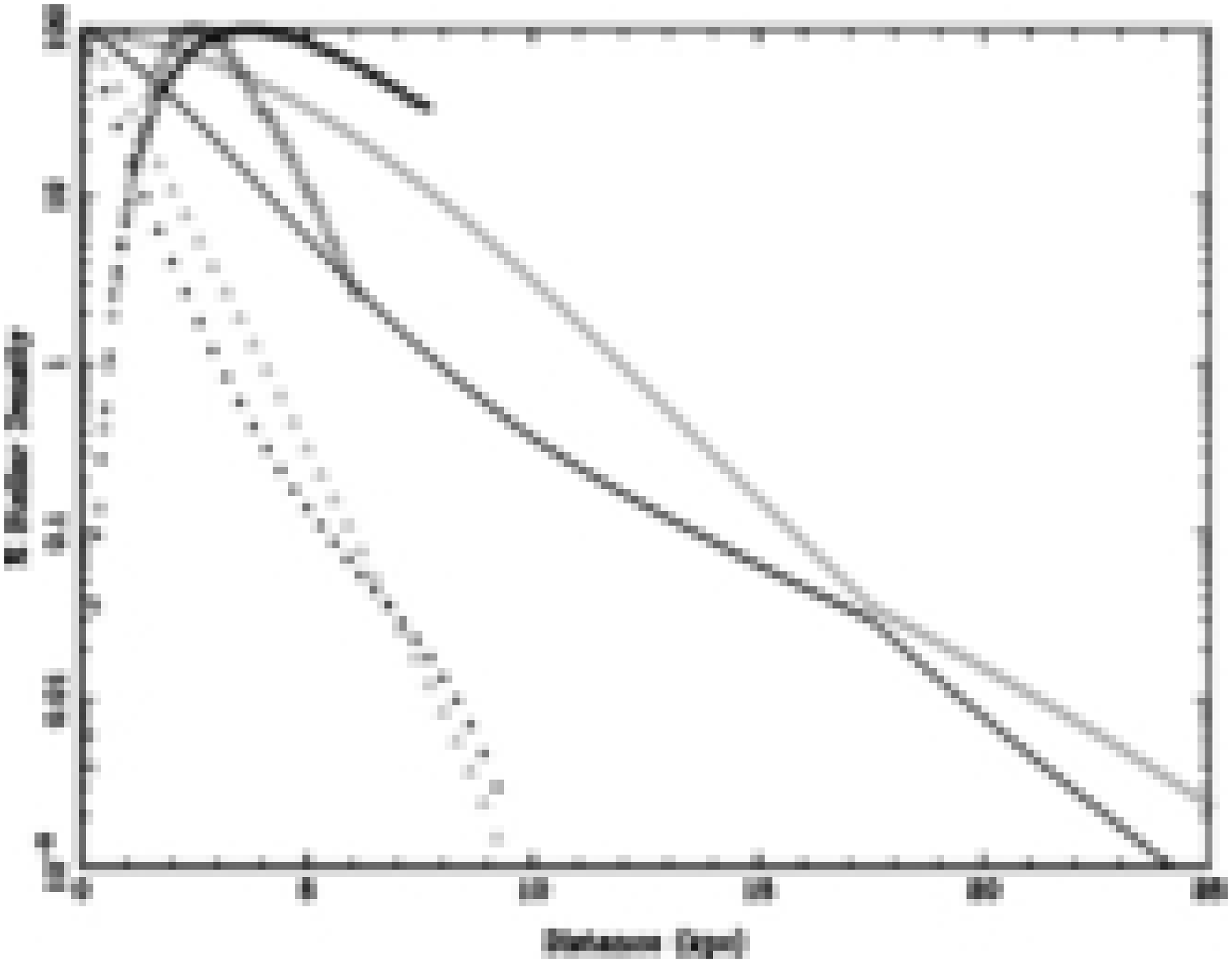}
\caption[]{\small Line-of-sight stellar density for ({\it l,b}) =
  (240,$-$9)$\deg$ and (240,$+$4)$\deg$ fields. The filled symbols are for
  the Southern field and the empty symbols for the Northern
  fields. The various models are the Besan\c{c}on synthetic Galaxy model
  (stars), \citet{Yusi2004} model (circles) and the
  \citet{2002A&A...394..883L} model (triangles).
\label{figthreemodel}}
\end{figure}
The existence of the Canis Major Dwarf is debated
by
\citet{2005ApJ...630L.153C}, \citet{2006MNRAS.368L..77M}, \citet{2006A&A...451..515M}
  and \citet{2006MNRAS.369.1911L},
who propose either a  new warp or spiral arm scenario to explain
the presence of the overdensity in this region. While all the issues
raised in these papers cannot be addressed here, some qualitative
comparisons can provide further insight into this debate.  

\subsubsection{Warp or Dwarf?}\label{warp}
The primary source of doubt over the presence of the dwarf galaxy is
whether a warp in the disc of the Galaxy is sufficient to explain the
overdensity of M giant stars as reported in
\citet{2004MNRAS.348...12M}.  The warp in the Galaxy has been known
for a long time and was primarily described through observations of
the gas disc, as presented in \citet{1988gera.book..295B}. Tracing the
warp through the stellar populations is more difficult, as only certain
types of stars have reliable distance estimate
techniques. \citet{Yusi2004} exploits the pulsar distribution around 
the Galaxy, while \citet{2002A&A...394..883L} uses the 2MASS catalogue
to trace the warp with the old stellar population (giant stars \& Red
clump stars). The basis of these studies is the assumption that the Galaxy is
essentially symmetric above and below the plane. Following the path of
the symmetry reveals the warp in the disc and its deviation from the
{\it b} = 0$\deg$ plane. \citet{2006A&A...451..515M} also employs the 2MASS
catalogue using the raw counts of Red Giant Branch and Red clump
stars above and below the plane to trace the warp. They argue their
result shows the mid-plane of the disc is shifted by $\sim$3$\deg$
below the plane in the direction of the Canis Major overdensity at
{\it l}$\sim$240$\deg$. This result does not elucidate how they account for the presence
of dust around the Galaxy and stars at all latitudes and reddening are
included. Although this presents a potential flaw in
their argument, we can nonetheless proceed to examine the data in
light of their claims. The survey presented here contains a strip of
observations at {\it l}$\sim$240$\deg$, which provide an excellent opportunity to try to
understand the stellar populations of this region. Given that all of
the warp models rely solely on symmetric stellar populations around
the plane to derive the warp, this can be used to interpret the CMDs
presented here. If we consider the field at ({\it l,b}) =
(240,$-$9)$\deg$ as our basis for the warp/dwarf problem, then, by
symmetry, there should be a field which is similar on the Northern side
of the plane. Figure~\ref{figthreehess} shows three CMDs from the
strip of observations at  {\it l}$\sim$240$\deg$ chosen to aid our
understanding. The field on the left is the observation at ({\it l,b})
= (240,$-$9)$\deg$, the centre is (240,$+$10)$\deg$ and the right CMD is
(240,$+$4)$\deg$. Overlaid on each of the CMDs is the fiducial sequence from \citet{2002ApJ...569..245N} used in the previous
sections. It is placed at a magnitude offset corresponding to $\sim$7.3 \kpc\ and serves as a reference point
for all the stars at that distance. If the Galaxy had no warp, then the
{\it b} = $-$9$\deg$ and {\it b} = $+$10$\deg$ fields should be, by symmetry,
almost identical.  Clearly, both the strength and location of the
dominant Main Sequence feature in the
South is unmatched in the North. This supports the presence of the
warp in this part of the Galaxy. \citet{2006A&A...451..515M} find the mid-plane
of the Galaxy to be offset by $\sim$3$\deg$ below the plane at {\it
l}$\sim$240$\deg$. This implies that the proper corresponding field to
compare the (240,$-$9)$\deg$ field is not (240,$+$10)$\deg$, but
rather (240,$+$3)$\deg$. Unfortunately, this survey does not include a
field in this location and so the (240,$+$4)$\deg$ is shown instead. The
(240,$+$2)$\deg$ could also be considered but is affected by too much
extinction to be useful in this comparison. Again, the Southern field
contains many more stars along the entire length of the dominant sequence
than its Northern counterpart, showing that the symmetry argument is
unsuitable for this field. Therefore, the warp is not an adequate explanation for the CMa overdensity. This simple test reveals that
these fields are not symmetric around the warped plane. The reasons for
supporting the presence of the dwarf galaxy is not to merely
substitute it for the warp, but rather to show that there are more stars in
this location than can simply be explained by adjusting the warp in
the Galactic disc. Comparing the predictions of each model for the fields ({\it
l,b}) = (240,$-$9)$\deg$ and (240,$+$4)$\deg$, Figure~\ref{figthreemodel} shows the stellar
density profiles as predicted by the \citet{Yusi2004} model, the
\citet{2002A&A...394..883L} model and the Besan\c{c}on model. The
Besan\c{c}on model's prediction is generated via the histogram of star
counts taken from the same data source as the comparison
fields generated in Figures~\ref{fig240m9} and \ref{fig240p4}. The
abrupt cut-off in the Besan\c{c}on model is from only selecting the
thin disc stars in this analysis. The other two models are generated via
the density equations presented in their respective
papers \citep[][]{2002A&A...394..883L,Yusi2004}. Figure~\ref{figthreemodel} clearly shows that none of the models
predict any rise in stellar density at the 7.3\kpc\ distance
of the CMa overdensity. Only the Besan\c{c}on model shows a rolling
over of the stellar density at $\sim$5\kpc\ for the Southern field and
$\sim$3\kpc\ for the Northern field. This does favour higher
stellar densities in the South but comparisons with the model, as in
Figure~\ref{fig240m9}, show that there is a clear discrepancy between
the model and the data. Although the
\citet{2002A&A...394..883L} model maintains roughly the same stellar
density profile in both hemispheres, at only $\sim$5 \% of the maximum
stellar density in that direction the density
is far too low at the distance of the dwarf to account for the number
of stars seen there. The \citet{Yusi2004} model with different warp
parameters predicts more stars in the North than in the South as seen
in Figure~\ref{figthreemodel} at {\it b} = +4$\deg$ and -9$\deg$.

The AAT/WFI data presented here does suggest that there is an
overdensity of stars below the plane at {\it l} = 240$\deg$, which is
still unexplained by the latest warp models. Although the Besan\c{c}on
model can be criticised for introducing a thin disc cut-off and not
having the latest estimations of the warp and flare, it does attempt to model the entire set of properties considered to be part of
the Galaxy. New parameters of the warp and flare need to be incorporated into
the entire picture of the Galaxy to be truly useful when
presented with actual data. These additional stars are unexpected in a
mostly symmetric Galaxy, but whether they belong to a dwarf
galaxy or not is difficult to tell. It is certain though that the
standard Galaxy model is inadequate and that a dwarf galaxy could
introduce a strong sequence as seen in the CMa region.

\subsubsection{The Blue Plume Star Problem}\label{BPstars}

\begin{figure}\centering
\includegraphics[width=7cm,angle=270]{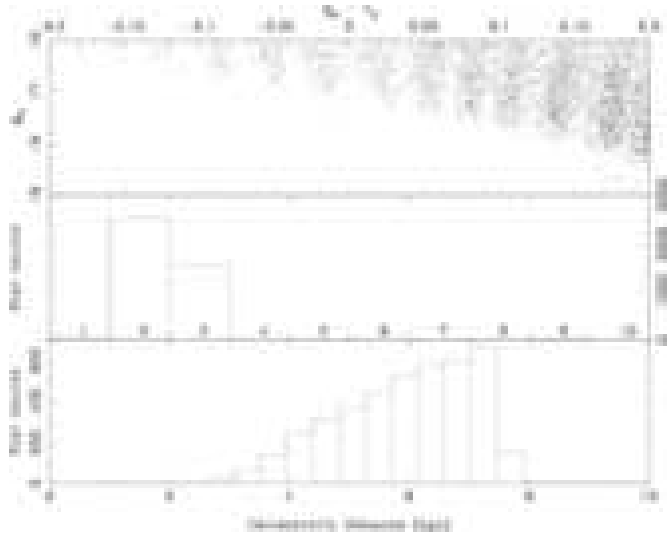}
\caption[]{\small Blue Plume stars selected from the Besan\c{c}on synthetic
  Galaxy model in the field (240,$-$4)$\deg$. The top panel shows their
  location on a Colour-Magnitude diagram. The middle panel is their
  distribution of Ages (parametrized into 10 bins), with Age 2
  corresponding to stars 0.15 - 1 Gyr old and Age 3 being stars 1 - 2
  Gyrs old. The lower panel is the heliocentric distance distribution
  of the stars in kpc. The striation in the top panel is due to the
  resolution of the model and is not an observable phenomena.
\label{figBPbesplotsouth}}
\end{figure}
\begin{figure}\centering
\includegraphics[width=7cm,angle=270]{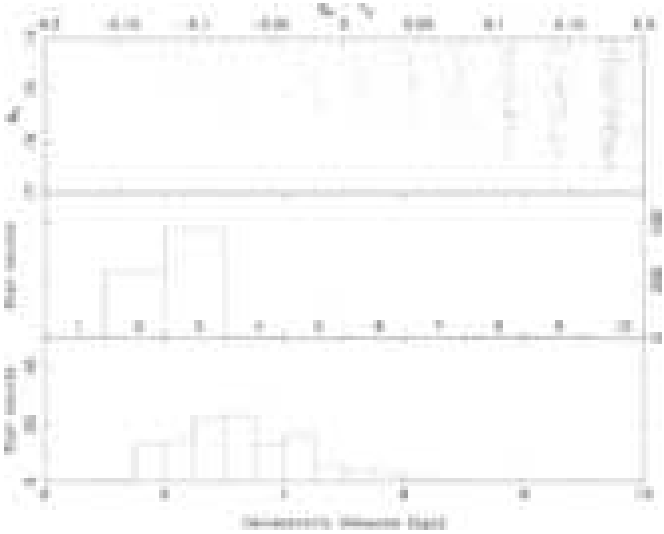}
\caption[]{\small Blue Plume stars selected from the Besan\c{c}on
  synthetic Galaxy model in the field (240,$+$4)$\deg$. The top panel
  shows their location on a Colour-Magnitude diagram. The middle panel
  is their distribution of Ages (parametrized into 10 bins), with Age
  2 corresponding to stars 0.15 - 1 Gyr old and Age 3 being stars 1 -
  2 Gyrs old. The lower panel is the heliocentric distance
  distribution of the stars in kpc. The striation in the top panel is
  due to the resolution of the model and is not an observable
  phenomena.
\label{figBPbesplotNorth}}
\end{figure}
Investigating the presence of the Blue Plume (BP) stars in the region of the CMa
overdensity has unveiled an interesting problem. The BP stars can be
seen in Figure~\ref{fig240m9} around 18$^{th}$ magnitude and in the colour
range ($g-r$) = 0.0 - 0.3, however around 0.3 there will be some
contamination from Main Sequence stars. Recently these stars have been
connected with the Disc, the CMa dwarf galaxy and a distant spiral
arm. So to which structure do the BP stars belong? The following section
will present the arguments for both sides and conclude with a
possible course of action to resolve this problem.

The evidence in favour of the dwarf galaxy has relied on the presence
of BP stars in the CMa CMDs. Figure~1 from \citet{2005ApJ...633..205M}
has illustrated that a model dwarf galaxy CMD could be consistent with these
stars being part of the dwarf galaxy. More recently \citet{butler2006}
has shown that the BP stars follow a different distribution, more
confined to the Galactic Plane, than the Red Clump stars which were
used to identify the CMa overdensity. However, as proposed by
\citet{2005ApJ...630L.153C} and \citet{2006MNRAS.368L..77M}, these
stars might be associated with a much more distant structure, an outer
spiral arm. The present survey, consisting of only two filters, is unable to
confirm the distance estimates of these previous studies, but using
the large range of longitudes available it can highlight where the BP
stars are visible.

Examination of the data reveals one possible solution to the BP star
problem. The BP stars are clearly visible in the two most important
fields of the CMa overdensity: (240,$-$9)$\deg$ as shown in
Figure~\ref{fig240m9} and (245,$-$9)$\deg$ in
Figure~\ref{fig245m9}. Close inspection of the other fields show
that in fact all the fields at {\it l} $\ge$ 240$\deg$ {\it below} the
plane have a BP population. Interestingly, none of the fields above
the plane show any evidence of BP stars and those fields below the
plane at {\it l} = (193 - 220)$\deg$ also have no BP stars. So the BP
stars in the present survey are visible solely below the plane and at {\it l}
$>$ 220$\deg$. Although, the fields {\it l} = (193 - 220)$\deg$ may be
too far out of the plane for the BP stars to be seen. This is not an
issue for the Northern fields, as the Disc is sampled at several
latitudes and the BP stars are not present. The only information
provided by the present survey on the BP stars is the magnitude at which they
are located on the CMD and the direction in which they were
observed. A direct measurement of their distance is not possible, but
since the distance to the sequences in the CMD can be estimated
via fitting of the fiducial, a qualitative approach to the BP distance
can be made. The first problem with the BP stars arises here. In each of the AAT/WFI fields containing BP stars, the
magnitude at which they are seen is almost constant. Indeed, while the
main components of the CMD become fainter with increasing longitude,
these stars maintain the same brightness levels. It can be concluded
then that these stars are not associated with the strong sequences seen in the CMDs and in turn are not associated with the
dwarf galaxy.

The Besan\c{c}on comparison fields of
(240,$-$6)$\deg$, (240,$-$4)$\deg$ and (240,$-$2)$\deg$
(Figures~\ref{fig240m6}, \ref{fig240m4} and \ref{fig240m2}) also
contain a BP population at a magnitude comparable with those
seen in the data; perhaps then this is not the detection of a new
structure, but rather an accepted component of the Galaxy.  A
breakdown of the BP stars in the Besan\c{c}on model
provides some insight into their origins (see the analysis of the
field at ($l,b$)$=$($240,-4$)$^\circ$ in
Figure~\ref{figBPbesplotsouth}). The ``Age'' of the star in
Figure~\ref{figBPbesplotsouth} indicates whether it belongs to the
Thin disc, Thick disc, Bulge or Halo. An Age of 1 - 7 corresponds to Thin disc
stars, Age 8 - Thick disc, Age 9 - Halo and Age 10 - Bulge. The
majority of the stars seen in the BP region are of Ages 2 or 3, which
is comparable to a population 0.15 - 1 Gyr old for Age 2
and 1 - 2 Gyr old for Age 3. While many BP stars are local, out of the
$\sim$4100 stars plotted here in this colour-magnitude range, the
distribution peaks at $\sim$7 \kpc\ and drops rapidly to zero by 8
\kpc. This is due to cutting through the Thin disc in this direction. The missing BP
stars in the Northern fields of the survey can be explained by the
Besan\c{c}on model shown in Figure~\ref{figBPbesplotNorth}. Here, at ({\it
l,b}) = (240,$+$4)$\deg$, the BP stars are found in a magnitude range
too bright to be observed at this location, as seen in the data from
the present survey. The number of BP stars in this
field is approximately a factor of 10 times fewer than those of its Southern
counterpart. The modelled Northern BP stars are also significantly closer
at only $\sim$3 \kpc. Both the paucity of BP stars and
their proximity lead to the conclusion that these stars belong to the local
Disc. With the Disc being warped, the line-of-sight for the Northern
field exits the Disc earlier, reducing the numbers of BP stars
visible. It also ensures the population appears closer; the reverse is
true for the South. So when taking into account the predictions of the
model, a plausible explanation for the BP stars being part of the Disc
is found. 

\subsubsection{Spiral Arms}
\begin{figure}\centering
\includegraphics[width=80mm,angle=270]{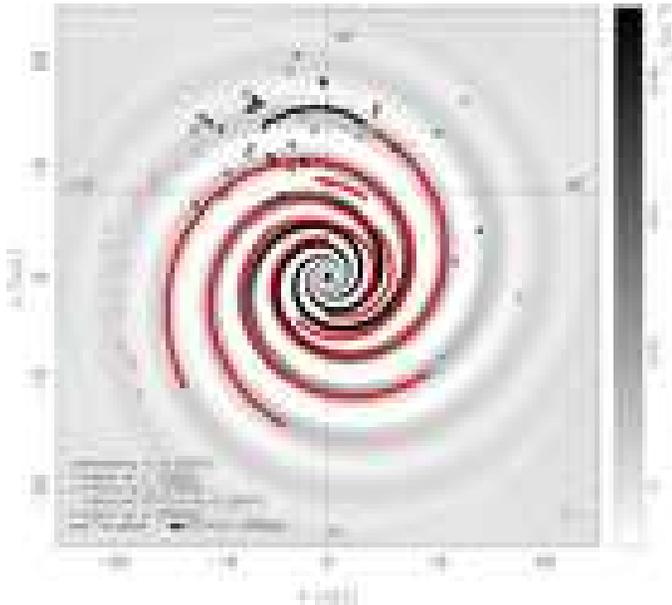}
\caption[]{\small Location of the Monoceros Ring and Canis Major detections
  shown with the grayscale map of the differential HI
  density for a simple four-arm Milky Way spiral model, the
  spiral model of \citet{cordes02} is overlaid as solid red (gray) lines. This is a
  reproduction of the left panel of Figure 3 from \citet{2004ApJ...607L.127M}, who
  kindly provided their Fortran code to allow our points to be overlaid on
  their figure. The filled symbols represent detections above the
  plane, with the empty symbols those below.  The MRi (above the plane) as discussed in
  \citet{2006ApJ...651L..29G} is shown as a solid black line. Note
  that this stream my appear as an extension of a Galactic spiral arm,
  but is, in fact, $\sim30^\circ$ above the Galactic plane. The crosses mark the location of the distant spiral
  arm detections as mentioned in \citet{2004ApJ...607L.127M}. The Sun
  is located at (0.0,8.0), with the lines intersecting the Sun marking
  out the Galactic longitudes of 0$\deg$, 90$\deg$, 180$\deg$ and
  270$\deg$. A $\pm$1 \kpc\ error bar in the lower right corner is
  indicative of the accuracy of these detections.
\label{figspiral}}
\end{figure}

It has been conjectured that both the MRi and CMa overdensities can be
regarded as part of normal Galactic structure [see
\citet{2005ApJ...630L.153C} and \citet{2006MNRAS.368L..77M}]. The MRi is considered
to be an extension of the flare into higher latitudes
\citep{2006A&A...451..515M} and the CMa overdensity a line-of-sight effect
close to the maximum Galactic warp \citep{2006MNRAS.369.1911L}. To
investigate how well the detections of both these structures align
with the spiral arms, we have overlaid their locations on Figure 3 of
\citet{2004ApJ...607L.127M} as shown in Figure~\ref{figspiral}. The
results of that paper have now been extended in \citet{levine2006}
revealing the presence of distant gaseous spiral arms out to $\sim$25
\kpc. Figure~\ref{figspiral} consists of a modelled grayscale density
map of the differential HI distribution with the spiral arm models of
\citet{cordes02}, over-plotted as solid red (gray) lines. On this, the locations of
the MRi and putative CMa detections have been included in the same
manner as Figure~\ref{figalldetect}. 

Do the locations of the MRi detections suggest that it is part of a
distant spiral arm? Most of the MRi detections do not seem to
correlate with any of the nearby spiral arm locations, with the
exception of the \citet{2006ApJ...651L..29G} which is the result of
the distance estimate of 8.9$\pm$0.2 \kpc, however, the
\citet{2006ApJ...651L..29G} portion of the stream outlined here is
found between {\it b} = (17 - 38)$\deg$, well above the plane. The
remaining points almost seem to align with the gaps between spiral arms more strongly than with the
spiral arms themselves. The fields in the first quadrant at {\it l} =
61$\deg$ and 75$\deg$ are both close to spiral arm overdensities.  The
difficulty with connecting them to spiral arms is that the Northern
fields reside at {\it b} = $+$15$\deg$, and both are at about 15 \kpc\
from the Sun. This places them around 3.7 \kpc\ out of the plane, well
outside the warped flared regions in this direction. The field below
the plane at ({\it l,b}) = (61,$-$15)$\deg$ is at 5.4 \kpc\ out of the
non-warped plane and so is further from the plane when considering the effect of the
Northern warp. The detection at ({\it l,b}) = (118,$+$16)$\deg$ is 3.3
\kpc\ out of the non-warped plane and is again beyond the warped
flared disc of the Galaxy. Continuing on around the Galactic plane, the fields at {\it l} =
123$\deg$ have been discussed in $\S$\ref{flare} and the remaining MRi
detections from the various other authors all typically lie between {\it b}
= (20 - 30)$\deg$ and cannot be associated with a warped and flared
Disc. For the MRi at least, any alignment with the spiral arms appears
purely coincidental, as the density profiles of the Disc do not allow
for such overdensities to reside well above the Plane.

Do the CMa locations follow the spiral arm? The CMa overdensity fields
can be seen as the nearby open star symbols from {\it l} =
(220 - 273)$\deg$ in Figure~\ref{figspiral}.  The CMa fields are
certainly close to the spiral arm, but are the progressively shorter
distance estimates to these detections a statistical
effect related to the distance estimation technique or does it
represent a real change in direction for the overdensity? If the field
at ({\it l,b}) = (193,$-$21)$\deg$ is part of the overdensity, then CMa
would be seen as an addition to the Disc component. This is because
the location of the dominant sequence in that field is further below
the plane than is expected. At $\sim$5.8 \kpc\ from the Sun, it is 2.1 \kpc\ below
the plane, much further than the 1.0 \kpc\ predicted by the
Besan\c{c}on synthetic Galaxy model. If this is not an isolated
overdensity of stars, then perhaps it is the CMa overdensity extended
to these longitudes.  Disentangling the CMa overdensity from the Disc
is not simple, but fields contained in the present survey suggest that a pure
warp scenario and hence a spiral arm theory cannot explain all of the
observations. Indeed spiral arms seem not to be an identifiable
feature within the CMDs. The INT/WFC fields at ({\it l,b}) =
(61,$\pm$15)$\deg$ cross 2 - 3 spiral arms and there is no
distinguishable sign of these features in the CMDs. The spiral arms
are not visible in the old Main Sequence stellar populations which are
used to characterise both the MRi and CMa detections. The overdensity
in CMa therefore cannot be considered a ``normal'' or additional
spiral arm. A more detailed study of this region is imperative to
resolve all these issues.

\subsubsection{The Canis Major Stream}\label{cmastream}
\begin{figure*}\centering
\includegraphics[width=140mm,angle=270]{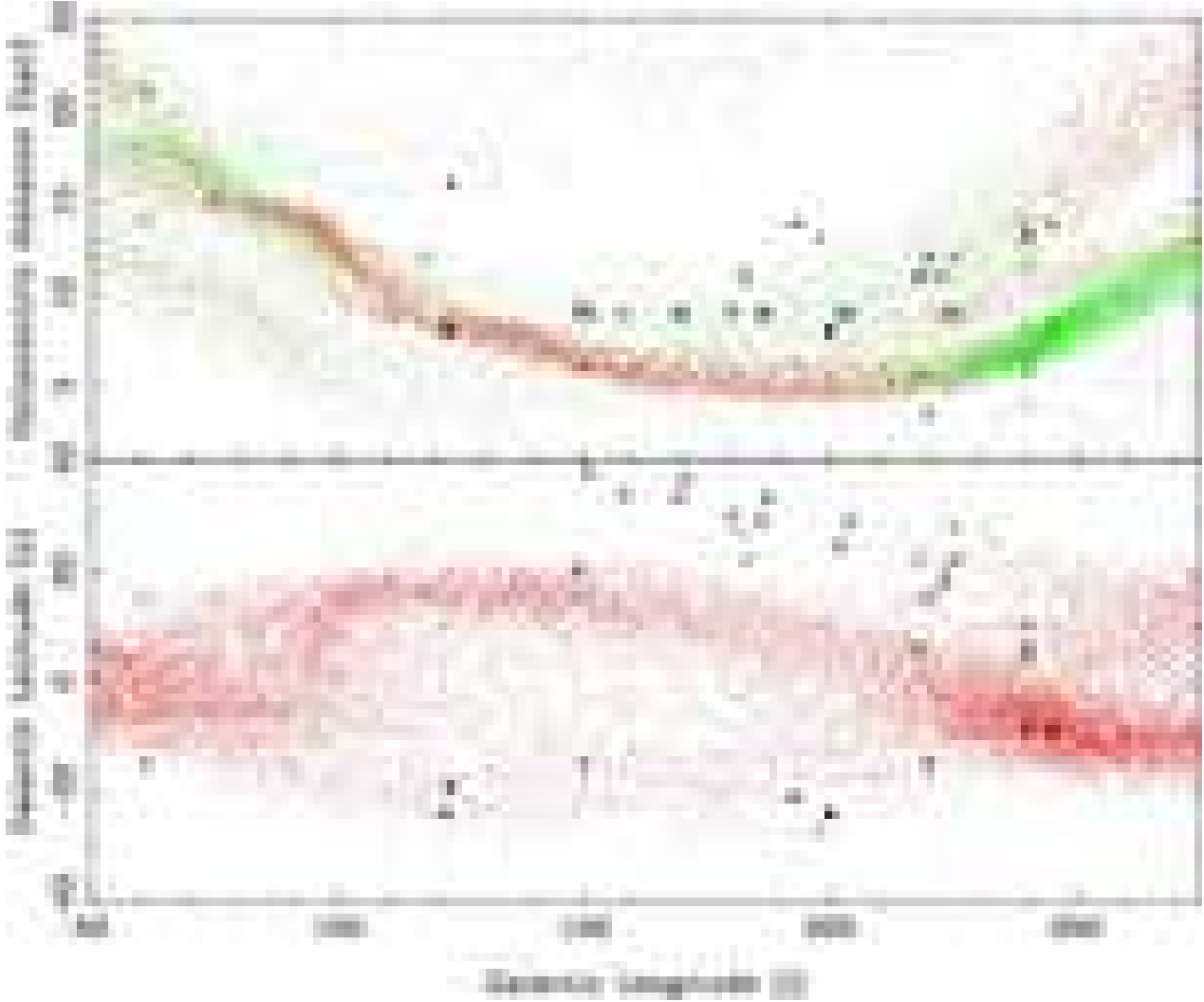}
\caption[]{\small Location of the Monoceros Ring and Canis Major detections
  shown with the numerical simulation of the Canis Major dwarf galaxy
  as presented in \citet{2005MNRAS.362..906M}. The symbols are as follows:
  in the top panel the simulation is shown with red (dark gray) dots representing
  stars below the plane and green (light gray) above the plane. Filled
  symbols represent observations below the plane and open symbols are
  observations above the plane. The circles are from
  \citet{2005MNRAS.362..475C}; the stars are from
  \citet{2003MNRAS.340L..21I}; the diamonds are from
  \citet{2002ApJ...569..245N}; the asterisks are from
  \citet{2003ApJ...588..824Y}; the squares are from the present paper with
  the plus (+) symbols for the CMa/Disc distance in the northern
  fields and the cross (x) symbols for the CMa/Disc distance in the
  southern field. The outline of the stream described in
  \citet{2006ApJ...651L..29G} is shown with the circles containing the
  plus signs.
\label{figCMastream}}
\end{figure*}

Given the interest in the Monoceros Stream and the subsequent proposal
of a dwarf galaxy in Canis Major, it seems obvious to ask where the
immediate tidal tails of the dwarf are to be found. The presence of an
obvious tidal feature associated with the overdensity would strengthen
the idea that it is a dwarf galaxy. To date, there have only
been two simulations of the MRi/CMa system. The simulation from
\citet{2005MNRAS.362..906M} uses kinematics of the CMa overdensity as
constraints on the model, while \citet{2005ApJ...626..128P}
constructs their simulation with data from the MRi. Although both
simulations predict the progenitor to be in the
Canis Major region, only the \citet{2005MNRAS.362..906M} simulation places it in
its currently accepted location. Given that this was the criteria for
the \citet{2005MNRAS.362..906M} model this result is
unsurprising but searching for the CMa stream with the
\citet{2005ApJ...626..128P} model is not feasible given it does
not coincide with current observations.  The following discussion will
focus on the distance estimates to both the MRi sequence in the
various fields and also to the distance of the dominant Main Sequence
in the CMDs. It is important to understand though that the dominant
sequence could be either a pure Disc population, a pure CMa
population or a mixture of both and although distinguishing between
them is not possible the result is compared with the
~\citet{2005MNRAS.362..906M} model.

Figure~\ref{figCMastream} shows the results of the numerical simulation of
\citet{2005MNRAS.362..906M} with the locations of the currently known detections of
both the MRi and CMa. This is an extension of Figure 20 as seen in
\citet{Bellazzini07}. Firstly, it should be noted that the CMa dwarf
galaxy is located at ({\it l,b}) $\sim$ (240,-9)$\deg$, as can be seen in the
overdensity of points in that location (lower panel) with the CMa stream
arcing into the northern hemisphere around {\it l} = 200$\deg$. The
top panel shows the heliocentric distance to the points, with those
below the plane in red and those above in green. CMa is
located at its accepted distance of $\sim$7 \kpc. The
other symbols are described in the Figure caption. A key feature of
the \citet{2005MNRAS.362..906M} and
\citet{2005ApJ...626..128P} models is that the MRi is simply a
wrapped tidal arm of the CMa accretion event. In this manner, the
structures described within \citet{2002ApJ...569..245N},
\citet{2003ApJ...588..824Y}, \citet{2003MNRAS.340L..21I} and
\citet{2006ApJ...651L..29G} conform to this idea of sampling a
wrapped tidal arm. Their detections are in general more distant and
less dense than the immediate CMa stream as predicted by the
\citet{2005MNRAS.362..906M} model. Some observations in both the present
paper and \citet{2005MNRAS.362..475C} do seem to coincide with the
predicted spatial location of the CMa stream on the
sky.\footnote{\citet{Bellazzini07} contains additional observations,
  not included here, which
also intersect with the predicted location of the CMa stream.} In
particular, the fields at ($l,b$) = (118,+16)$\deg$, (150,+15)$\deg$,
(218,+6)$\deg$, (260,-10)$\deg$ and (273,-9)$\deg$ are located within
the modelled CMa stream as it is seen on the sky (lower panel). Indeed, when
considering the distance estimates to the features in those fields
there is a good correspondence with the predictions of the model. In
the fields at ($l,b$) = (118,+16)$\deg$ and (150,+15)$\deg$, the
features are only a few \kpc\
more distant than the model but it is only constrained in the CMa
region, so this difference is not unexpected. At ($l,b$) = (218,+6)$\deg$, the
distance estimate to the CMa/Disc feature is consistent with the
model predictions, as is also the case for the ($l,b$) = (260,-10)$\deg$ field; the
($l,b$) = (273,-9)$\deg$ detection is found on the edge of the predicted distance estimate
of the stream. For those fields in the CMa region, an interesting
interpretation can be made. The detections below the plane (crosses)
seem to demarcate the far edge of the dwarf, while the detections
above the plane (plus signs) demarcate the near side. This can be seen
in the location of these detections in the top panel. Could this be inferring
the orientation of the dwarf?  

Although this rudimentary coincidence of the observations of the
MRi/CMa fields with the \citet{2005MNRAS.362..906M} model does not
validate this model. It does suggest that previous observations termed
MRi detections could be reinterpreted as CMa stream detections. In particular, both
the ($l,b$) = (118,+16)$\deg$ and (150,+15)$\deg$ fields from
\citet{2005MNRAS.362..475C} would fit this new scenario. The
fields from the present paper which align with the CMa stream also
support this conclusion. Unfortunately, there are too few
observations between {\it l} = (100 - 180)$\deg$ to further test the
model in these regions. With the CMa stream predicted to be relatively
nearby, testing these predictions at latitudes around {\it b} = 20$\deg$ should
be fairly straightforward. 

\section{Conclusion}\label{conclusion}
The survey presented here consists of 16 fields from {\it l} = (193 - 276)$\deg$, all
observed with the AAT/WFI between 2004 and 2006, providing strong evidence that the Monoceros Ring cannot be
considered part of the normal warp/flare profile of the Galactic
disc. Of the 16 fields, 8 have a strong sequence
beyond the normal Disc component which have been interpreted as the
Monoceros Ring; two others are presented as tentative
discoveries. Resolving the origins of the putative Canis Major dwarf
galaxy is
extremely difficult with a survey of this kind, but the conclusions
reached are that by symmetry around the warped Galactic plane,
there is no field above the plane which matches the properties of
those in the overdensity. Also, there is no appreciable change in the
density profiles of the various Galactic disc models to explain an
overdensity of stars at that distance in Canis Major. The origins of
the Blue Plume stars reveal two contradictory scenarios. From one view
they are part of the Galactic disc and from another they are not. Although a
more detailed survey is needed to resolve this issue, they can no longer be associated with Canis Major
overdensity stars. Searching for a spiral arm explanation to
the Monoceros Ring and Canis Major detections are mostly excluded on
the basis of their distance out of the plane. The Canis
Major detections are close to the plane and partially mixed in with known
Galactic components. This makes disentangling them very
difficult. However, since spiral arms are not visible in the basic
structure of CMDs, which are comprised mainly of old Main Sequence stars,
attributing the putative dwarf to an outer spiral arm is not possible with the data in hand. Importantly,
the location of the dominant sequence in the ({\it l,b}) =
(193,$-$21)$\deg$ field is highly inconsistent
with current Galaxy models and may represent and extension of the CMa
overdensity into this field. This field differs the most from the Besan\c{c}on synthetic Galaxy model and may indicate a
location where the CMa dwarf can be analysed away from the Disc. A
study to fill the gaps in the entire CMa region is required to
determine the origins of these disputed structures. Furthermore, the
existence of a CMa stream can now be considered a possibility with
previous detections labelled as MRi detections now potentially associated with the
CMa stream. Further sampling of the predicted locations of the CMa
stream is necessary to resolve the uncertainties presented
here. Although many properties of the MRi and the CMa/Disc features
are unknown, there is now a growing pool of evidence supporting a
merger event in the Galaxy. Continued study will undoubtedly reveal
their true impact on the formation and evolutionary scenarios of the Milky Way.

\section{Acknowledgements}
Many thanks to the referee for their thorough review which has
improved the quality of this paper. We would also like to thank Naomi
McClure-Griffiths for her assistance with the spiral arm
comparisons. BCC would like to thank the University of Sydney for its
UPA Scholarship and ESO for their Postdoctoral Fellowship for without
them he would not be able to listen to the cricket from a telescope on
the other side of the world. RRL thanks both LKN and MK for their
financial and on-going support. GFL acknowledges the support of the
Discovery Project grant DP0343508.  The research of AMNF has been
supported by a Marie Curie Fellowship of the European Community under
contract number HPMF-CT-2002-01758.

\newcommand{\aap}{A\&A}
\newcommand{\apj}{ApJ}
\newcommand{\apjl}{ApJ}
\newcommand{\aaps}{AAPS}
\newcommand{\aj}{AJ}
\newcommand{\mnras}{MNRAS}
\newcommand{\nat}{Nature}

\bsp

\label{lastpage}

\end{document}